\documentclass[twocolumn]{svjour3}          

\usepackage{times}
\usepackage{epsfig}
\usepackage{graphicx}
\usepackage{amsmath}
\usepackage{amssymb}
\usepackage{multirow}
\usepackage{lscape}
\usepackage{longtable}
\usepackage{adjustbox}
\usepackage{enumitem}
\usepackage{array}
\usepackage{colortbl}
\usepackage{hhline, soul}
\usepackage[ruled,vlined,linesnumbered]{algorithm2e}
\usepackage{makecell}
\usepackage{acronym}
\usepackage{cellspace}
\usepackage{booktabs}
\usepackage{nccmath}
\usepackage[table,xcdraw]{xcolor}
\usepackage[pagebackref=true,breaklinks=true,colorlinks,bookmarks=false]{hyperref}
\newacro{dl}[DL]{Deep learning}
\newacro{ae}[AE]{Adversarial Example}
\newacro{sfad}[SFAD]{Selective and Feature based Adversarial Detection}
\newacro{dnn}[DNN]{Deep Neural Network}
\newacro{cnn}[CNN]{Convolutional Neural Network}
\newacro{nn}[NN]{Neural Network}
\newacro{hca}[HCA]{High Confidence Attack}
\newacro{pgd}[PGD]{Projected Gradient Descent}
\newacro{lid}[LID]{Local Intrinsic Dimensionality}
\newacro{fgsm}[FGSM]{Fast Gradient Sign Attack}
\newacro{nic}[NIC]{Neural-network Invariant Checking}
\newacro{gan}[GAN]{Generative Adversarial Networks}
\newacro{dnr}[DNR]{Deep Neural Rejection}
\newacro{mtl}[MTL]{Multi-Task Learning} 
\newacro{rbf}[RBF]{Radial Basis Function} 
\newacro{svm}[SVM]{Support Vector Machine}
\newacro{fs}[FS]{Feature Squeezing}
\newacro{ta}[TA]{Threshold Attack}
\newacro{pa}[PA]{Pixel Attack}
\newacro{st}[ST]{Spatial Transformation attack}
\newacro{cw}[CW]{Carlini-Wagner}
\newacro{df}[DF]{DeepFool}
\newacro{bu}[BU]{Bayesian Uncertainty}
\newacro{nss}[NSS]{Natural Scene Statistics}
\newacro{kd}[KD]{Kernel Density}
\newacro{sa}[SA]{Square Attack}
\newacro{eot}[EOT]{Expectation Over Transformation}
\newacro{bpda}[BPDA]{Backward Pass Differentiable Approximation}

\newcolumntype{P}[1]{>{\centering\arraybackslash\hspace{-\tabcolsep}}p{#1}}
\newcolumntype{M}[1]{>{\centering\arraybackslash}m{#1}}
\DeclareMathOperator*{\argmax}{argmax}
\newcommand{\etal}{\textit{et al. }}
\newcommand{\hop}{HopSkipJump }
\definecolor{customgreen}{RGB}{24, 186, 24}

\begin{document}
\title{Revisiting Model's Uncertainty and Confidences for Adversarial Example Detection
}


\author{Ahmed Aldahdooh         \and
        Wassim Hamidouche       \and
        Olivier D\'eforges        
}


\institute{All authors \at Univ Rennes, INSA Rennes, CNRS, IETR - UMR 6164, F-35000 Rennes, France\\
            \and
            Ahmed Aldahdooh -  
              \email{ahmed.aldahdooh@insa-rennes.fr}      \\     
           \and
           Wassim Hamidouche - 
              \email{wassim.hamidouche@insa-rennes.fr} \\
          \and
          Olivier Deforges - 
                \email{olivier.deforges@insa-rennes.fr}
}

\date{Received: date / Accepted: date}
\maketitle

\begin{abstract}
  Security-sensitive applications that rely on \acp{dnn} are vulnerable to small perturbations that are crafted to generate \acp{ae}. The \acp{ae} are imperceptible to humans and cause \acs{dnn} to misclassify them. Many defense and detection techniques have been proposed. Model's confidences and Dropout, as a popular way to estimate the model's uncertainty, have been used for \ac{ae} detection but they showed limited success against black- and gray-box attacks. Moreover, the state-of-the-art detection techniques have been designed for specific attacks or broken by others, need knowledge about the attacks, are not consistent, increase model parameters overhead, are time-consuming, or have latency in inference time.  To trade off these factors, we revisit the model's uncertainty and confidences and propose a novel unsupervised ensemble \ac{ae} detection mechanism that 1) uses the uncertainty method called SelectiveNet, 2) processes model layers outputs, i.e. feature maps, to generate new confidence probabilities. The detection method is called \ac{sfad}. Experimental results show that the proposed approach achieves better performance against black- and gray-box attacks than the state-of-the-art methods and achieves comparable performance against white-box attacks. Moreover, results show that \ac{sfad} is fully robust against \acp{hca} for MNIST and partially robust for CIFAR10 datasets. \footnote{The source code is available in \url{https://aldahdooh.github.io/SFAD/}}.\\
\end{abstract}

\section{Introduction} \label{sec:intro}
\ac{dl} has achieved remarkable advances in different fields in human life especially computer vision tasks like object detection, image classification \cite{krizhevsky2012imagenet,simonyan2014very,ren2015faster}, surveillance \cite{lecun2015deep}, and medical imaging \cite{shen2017deep}. Despite that, it is found that \ac{dl} models are vulnerable to adversaries \cite{szegedy2013intriguing,goodfellow2014explaining}. In image classification models, for instance, adversaries can generate \acp{ae}, by adding small perturbations to an input image that are imperceptible to humans and devices, that cause \ac{dl} models to misclassify the input images. Such potential threat affects security-critical \ac{dl}-based applications \cite{guo2018lemna} such as self-driving cars. 

\begin{figure*}[!ht]
\begin{center}
\includegraphics[width=\textwidth, keepaspectratio]{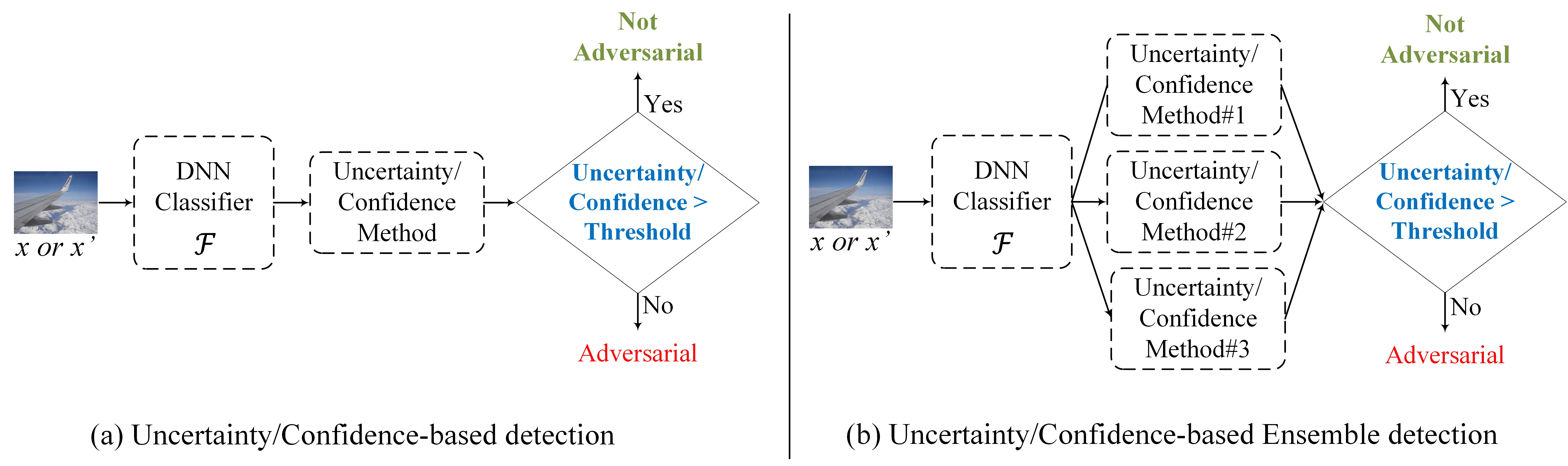}
\end{center}
\caption{(a) High-level architecture of the uncertainty/confidence-based detectors. (b) High-level architecture of the uncertainty/confidence-based ensemble detectors. The input sample is passed to the \ac{cnn} model to do class prediction. The detector, i.e. the uncertainty method, estimates the uncertainty of the input samples using model hidden layers. Using a predefined threshold, the input sample is not adversarial if the uncertainty exceeds the predefined threshold.}
\label{fig:abstract_design}
\end{figure*} 

Adversaries can generate \acp{ae} for white-box, black-box, and gray-box attacks \cite{akhtar2018threat,hao2020adversarial}. In white-box attack scenario, the adversary knows everything about the \ac{dl}-model including inputs, outputs, architecture, and weights of the model. Hence, he is guided by the model gradient to generate \ac{ae} by solving an optimization problem \cite{goodfellow2014explaining,kurakin2016adversarial,moosavi2016deepfool,carlini2017towards,madry2017towards}. In black-box scenario, the adversary knows nothing about the model but he leverages the transferability property \cite{papernot2016transferability} of \acp{ae} and the input content. By sending queries to the model, the adversary can craft small perturbations that are harmonious with the input image \cite{chen2017zoo,engstrom2019exploring,su2019one,kotyan2019adversarial}. In the gray-box scenario, the adversary knows only the input and the output of the model and hence, he tries to substitute the original model with an approximated model and then uses its gradient as in white-box scenario to generate \acp{ae}. 

Researchers pay attention to this threat and several emerging methods have been proposed to detect or to defend against \acp{ae}. More details about defense and detection methods can be found in Section \ref{sec:rw}.

\ac{dl} model's uncertainty is one of the main methods that has been used to determine whether an input sample belongs to the training manifold. The uncertainty is usually measured by adding randomness to the model using Dropout technique\cite{gal2016dropout,srivastava2014dropout}. It is found that clean sample predictions do not change, when randomness is added, while it changes for \acp{ae}. Feinman \etal \cite{feinman2017detecting} proposed \ac{bu} metric that used Monte Carlo dropout to estimate the uncertainty to detect \acp{ae} that are near the classes manifold, while Smith \etal \cite{smith2018understanding,sheikholeslami2019minimum} used mutual information method to estimate the uncertainty. The prediction risk of these methods is higher compared to the recent uncertainty method, SelectiveNet \cite{selective2019}, that is used in this work. 
On the other hand, it was shown in  \cite{hendrycks2016baseline} that predicted class probabilities, i.e. model's confidence, of in-of-distribution samples are higher than of out-of-distribution. Model's confidence was used in  \cite{hendrycks2016baseline,aigrain2019detecting,monteiro2019generalizable,sotgiu2020deep} to implement \ac{ae} detectors. Uncertainty and confidence based detectors showed limited success against black- and gray-box attacks. Uncertainty and confidence based detectors are usually threshold-based detectors as shown in Figure \ref{fig:abstract_design}(a). To enhance detectors' performance, one recommendation goes to the direction of providing ensemble detection methods, as shown in Figure \ref{fig:abstract_design}(b). Although state-of-the-art detectors achieve promising results, they may have one or more limitation(s); not performing well with some known attacks \cite{xu2017feature}, broken by attackers \cite{athalye2018obfuscated,carlini2017magnet}, performance of baseline detectors is not consistent \cite{bulusu2020anomalous}, increase the model parameters overhead \cite{lust2020gran}, time consuming \cite{ma2019nic}, or introduce latency \cite{gao2020backdoor} in the inference time.

In this paper and in order to mitigate the aforementioned limitations, we revisit the model's uncertainty and confidence to propose a novel ensemble \ac{ae} detector that hasn't had any knowledge of \acp{ae}, i.e. unsupervised detector, as shown in Figure \ref{fig:detector_design}. The proposed method has the following attributes;  1) it investigates SelectiveNet capability in detecting adversarial examples since it measures the uncertainty with less risk. According to the author's knowledge, the SlelectiveNet \cite{selective2019} is not used in adversarial attacks detection models. 2) Unlike other detectors \cite{melis2017deep,sotgiu2020deep,lu2017safetynet}, the proposed method uses the model's last $N$-layers outputs, i.e. feature maps, to build $N$-\acp{cnn} $\mathcal{M}$ that have different processing blocks like up/down sampling, auto-encoders \cite{wang2017residual,liu2019end}, noise addition \cite{lecuyer2019certified,liu2018towards,liu2020does}, and bottleneck layer addition \cite{he2016deep} that make the representative data of last layers more unique to the input data distribution to yield better model's confidence. To reduce the effect of white-box attacks, the output of $\mathcal{M}$ is transferred/distilled to build the last \ac{cnn} $\mathcal{S}$. 3) The proposed model ensembles the proposed detection techniques to provide the final detector. This step has a great impact in reducing the adversary's capability to craft perturbations that can fool the detector, since he has to fool every detection technique. We name the proposed method as \acf{sfad}. The high-level architecture of the \ac{sfad} is illustrated in Figure \ref{fig:abstract_design}.

A prototype of \ac{sfad} is tested under white-box, black-box and gray-box attacks on MNIST \cite{lecun1998gradient}, and CIFAR10 \cite{krizhevsky2009learning}. Under the white-box attacks, the experimental results show that \ac{sfad} can detect \acp{ae} at least with accuracy of 89.8\% (many with 99\%) for all tested attacks except for the \ac{pgd} attack \cite{madry2017towards} with at least 65\% detection accuracy in average. For black- and gray-box attacks, \ac{sfad} shows better performance than other tested detectors. Finally, \ac{sfad} is tested under the \ac{hca} \cite{carlini2017adversarial} and it shows that it is fully(100\%) and partially(57.76\%) robust on  MNIST and CIFAR10 respectively. \ac{sfad} sets the thresholds to reject 10\% of clean images. Moreover, comparisons with state-of-the-art methods are presented. Hence, our key contributions are:
\begin{itemize}[noitemsep,topsep=8pt,itemsep=4pt,partopsep=4pt, parsep=4pt]
    \item We propose a novel unsupervised ensemble model for \ac{ae} detection. Ensemble detection makes \ac{sfad} more robust against white-box and adaptive attacks. 
    \item We investigate the SelectiveNet's, as an uncertainty model,  capability in detecting \acp{ae}. 
    \item We show that, by processing the feature maps of last $N$-layers, we can build classifiers for better confidence distribution. We provide an ablation experiments to study the impact of the feature processing blocks.
    \item \ac{sfad} prototype proves the concept of the approach and lets the door open in future to find the best $N$ layers and the best $N$(or $M$) \acp{cnn} combinations to build the detector's classifiers. 
    \item Unlike tested state-of-the-art detectors, \ac{sfad} prototype shows better performance under gray- and black-box attacks. \ac{sfad} prototype shows that it is fully robust on MNIST and partially robust on CIFAR10 when attacked with \acp{hca}. For instance, \ac{lid} method \cite{ma2018characterizing} reported very high detection accuracy on the tested attacks, but fails on \acp{hca} \cite{athalye2018obfuscated,carlini2017adversarial}.
\end{itemize}

\section{Related work}\label{sec:rw}

\subsection{Detection methods} \label{sec:rw_detect}
Defense techniques like adversarial training \cite{goodfellow2014explaining,madry2017towards,xie2020smooth,tramer2017ensemble}, feature denoising \cite{xie2019feature,borkar2020defending,liao2018defense}, pre-processing \cite{mustafa2019image,prakash2018deflecting}, and gradient masking \cite{papernot2016distillation,papernot2017practical,gu2014towards,nayebi2017biologically} try to make the model robust against the attacks and let the model correctly classify the \acp{ae}. On the other hand, detection methods provide adversarial status for the  input image. Detection techniques can be classified according to the presence of \acp{ae} in the detector learning process into supervised and unsupervised techniques \cite{bulusu2020anomalous}. In supervised detection, detectors include \acp{ae} in the learning process. Many approaches exist in the literature. In the feature-based approach \cite{nguyen2015deep,grosse2017statistical,metzen2017detecting,lu2017safetynet}, detectors use clean and \acp{ae} inputs to built their classifier models from scratch by using raw image data or by using the representative layers' outputs of a \ac{dnn} model. For instance, in \cite{lu2017safetynet}, the detector quantizes the last ReLU activation layer of the model and builds a binary \ac{svm} with \ac{rbf} classifier. As reported in \cite{lu2017safetynet}, this detector is not robust enough and is not tested against strong attacks like \ac{cw} attacks. While the work in \cite{grosse2017statistical} added a new adversarial class to the \ac{nn} model and train the model from scratch with clean and adversarial inputs. This architecture reduces the model accuracy  \cite{grosse2017statistical}. In the statistical-based approach \cite{feinman2017detecting,ma2018characterizing}, detectors perform statistical measurement to define the separation between clean and adversarial inputs. In \cite{feinman2017detecting}, \ac{kd} estimation, \ac{bu}, or combined models are introduced. Kernel-density feature is extracted from clean and \acp{ae} in order to identify \acp{ae} that are far away from data manifold while Bayesian uncertainty feature identifies the \acp{ae} that lie in low-confidence regions of the input space. \ac{lid} method is introduced in \cite{ma2018characterizing} as a distance distribution of the input sample to its neighbors to assess the space-filling capability of the region surrounding that input sample. The works in \cite{carlini2017adversarial,athalye2018obfuscated} showed that these methods can be broken. Finally, the network invariant approach \cite{metzen2017detecting,eniser2020raid} learns the differences in neuron activation values between clean input samples and \acp{ae} to build a binary \ac{nn} detector. The main limitation of this approach is that it requires prior knowledge about the attacks and hence it might not be robust against new or unknown attacks.

On the other hand, in unsupervised detection, detectors are trained with clean images only to identify the \acp{ae}. It is also known as prediction inconsistency models since it depends on the fact that \acp{ae} might not fool every \ac{nn} model. That's because the input feature space is almost limited and the adversary always takes that as an advantage to generate the \acp{ae}. Hence, unsupervised detectors try to reduce this limited input feature space available to adversaries. Many approaches have been presented in the literature. The \ac{fs} approach \cite{xu2017feature} measures the distance between the predictions of the input and the same input after squeezing. The input will be adversarial if the distance exceeds a threshold. The work in \cite{xu2017feature} squeezes out unnecessary input features by reducing the color bit depth of each pixel and by spatial smoothing of adversarial inputs. As reported in \cite{xu2017feature}, \ac{fs} is not performing  well  with  some  known  attacks like \ac{fgsm}. Instead of squeezing, denoising based approach, like MagNet \cite{meng2017magnet}, measures the distances between the predictions of input samples and denoised/filtered input samples. It was found in \cite{carlini2017magnet,liao2018defense} that MagNet can be broken and do not scale to large images. Recently, a network invariant approach was introduced \cite{ma2019nic}. They proposed a \ac{nic} method that builds a set of models for individual layers to describe the provenance and the activation value distribution channels. It was observed that \acp{ae} affect these channels. The provenance channel describes the instability of activated neurons set in the next layer when small changes are present in the input sample while the activation value distribution channel describes the changes with the activation values of a layer. The reported performance of this method showed its superiority against other state-of-the-art models but other works reported that the baseline \ac{nic}'s detectors are not consistent \cite{bulusu2020anomalous}, increase  model  parameters overhead \cite{lust2020gran}, are time consuming \cite{ma2019nic}, and increase the latency in the inference time \cite{gao2020backdoor}.

\textbf{Uncertainty-based detectors}. Following the observation that the prediction of clean image remains correct with many dropouts, while the prediction of \ac{ae} changes. Feinman \etal \cite{feinman2017detecting} proposed \ac{bu} metric. \ac{bu} uses Monte Carlo dropout to estimate the uncertainty, to detect those \acp{ae} that are near the classes manifold, while Smith \etal \cite{smith2018understanding} used mutual information method for such a task. In \cite{sheikholeslami2019minimum}, Sheikholeslami \etal proposed an unsupervised detection method that provides a layer-wise minimum variance solver to estimate model's uncertainty for in-distribution training data. Then, a mutual information based threshold is identified. 

\textbf{Confidence-based detectors}. Aigrain \etal \cite{aigrain2019detecting} built a simple \ac{nn} detector that uses the model's logits of clean and \acp{ae} to build a binary classifier. Inspired by the hypothesis of that, for a given perturbed image, different models yield different confidences,  Monteiro \etal \cite{monteiro2019generalizable} proposed a bi-model mismatch detection method. The detector is a binary \ac{rbf}-\ac{svm} classifier that takes as input the output of two classifiers of clean and \acp{ae}. On the other hand, Sotgiu \etal proposed an unsupervised detection method that uses the last $N$ representative layers' outputs of the classifier to built three \ac{svm} classifiers with \ac{rbf} kernel. The confidence probabilities of the \acp{svm} are combined to build the last \ac{svm}-\ac{rbf} classifier. Then, a threshold is identified to reject inputs that have less maximum confidence probability.

\subsection{SelectiveNet as an uncertainty model} \label{sec:rw_selectiv}
Let $\mathcal{X}$ be an input space, e.g. images, and $\mathcal{Y}$ a label space. Let $\mathbb{P}(X,Y)$ be the data distribution over $\mathcal{X} \times \mathcal{Y}$. A model, $f:X \rightarrow Y$, is called a prediction function, $ \ell : Y \times Y \rightarrow \mathbb{R}^2$ is a given loss function. Given a labeled set $S_k = {(x_i, y_i)}_{i=1}^{k}  \subseteq (\mathcal{X} \times \mathcal{Y})^k$ sampled i.i.d. from $\mathbb{P}(X,Y)$, where $k$ is the number of training samples. The true risk of the prediction function $f$ w.r.t. $\mathbb{P}$ is $R(f) \triangleq \mathbb{E}_{\mathbb{P}(X,Y)}[\ell (f(x),y)]$ while the empirical risk of the prediction function $f$ is $\hat{r}(f\mid S_k) \triangleq \frac{1}{k} \sum_{i=1}^{k} \ell (f(x_i),y_i)$.

Here, we briefly demonstrate the SelectiveNet as stated in \cite{selective2019}. The selective model is a pair $(f,g)$, where $f$ is a prediction function, and $g : \mathcal{X} \rightarrow \{0,1\}$ is a binary selection function for $f$,
\begin{ceqn}
\begin{equation}
    (f,g)(x) \triangleq \begin{cases} f(x), & \text{if } g(x) = 1; \\ \text{don't know}, & \text{if } g(x) = 0. \end{cases}
\end{equation}
\end{ceqn}

A soft selection function can also be considered, where $g : \mathcal{X} \rightarrow [0,1]$, hence, the value of $(f,g)(x)$ is calculated with the help of a threshold $\tau$ as expressed in the following equation

\begin{ceqn}
\begin{equation}
    (f,g)(x) \triangleq \begin{cases} f(x), & \text{if } g(x) \geq \tau; \\ \text{don't know}, & \text{if } g(x) < \tau. \end{cases}
\end{equation}
\end{ceqn}

The performance of a selective model is calculated using coverage and risk. The true coverage is defined to be the probability mass of the non-rejected region in $\mathcal{X}$ and calculated as 
\begin{ceqn}
\begin{equation}
    \phi(g) \triangleq E_{P}[g(x)],
\end{equation}
\end{ceqn}
while the empirical coverage is calculated as

\begin{ceqn}
\begin{equation}
    \hat{\phi}(g\mid S_k) \triangleq \frac{1}{k} \sum_{i=1}^{k}g(x_i)
\end{equation}
\end{ceqn}

The true selective risk of $(f,g)$ is
\begin{ceqn}
\begin{equation}
    R(f,g) \triangleq \frac{E_P[\ell (f(x),y)g(x)]}{\phi(g)},
\end{equation}
\end{ceqn}
while the empirical selective risk is calculated for any given
labeled set $S_k$ as 

\begin{ceqn}
\begin{equation}
    \hat{r}(f,g\mid S_k) \triangleq \frac{\frac{1}{k}\sum_{i=1}^{k}\ell (f(x_i),y_i)g(x_i)}{\hat{\phi}(g\mid S_k)}.
\end{equation}
\end{ceqn}

Finally, for a given coverage rate $0 < c \leq 1$ and $\Theta$, a set of parameters for a given deep network architecture for $f$ and $g$, the optimization problem of the selective model is expressed as:
\begin{ceqn}
\begin{align}
    &\begin{aligned}
        \theta^\ast = \operatorname*{arg\,min}_{\theta \in \Theta} (R(f_\theta,{g}_\theta))\\
        \textit{s.t. } \phi({g}_\theta) \geq c,
    \end{aligned}
\end{align}
\end{ceqn}

and can be solved using the Interior Point Method (IPM) \cite{potra2000interior} to enforce the coverage constraint. That yields to unconstrained loss objective function over samples in  $S_k$,

\begin{ceqn}
\begin{align}\
    &\begin{aligned}
        \mathcal{L}_{(f,g)} \triangleq \hat{r}_{\ell}(f,g\mid S_k) + \lambda \Psi(c-\hat{\phi}(g\mid S_k))\\
        \Psi(a) \triangleq \max(0,a)^2,
    \end{aligned}
\end{align}
\end{ceqn}

where $c$ is the target coverage, $\lambda$ is a hyper-parameter controlling the relative importance of the constraint, and $\Psi$ is a
quadratic penalty function. As a result, SelectiveNet is a selective model $(f, g)$ that optimizes both $f(x)$ and $g(x)$ in a single model in a multi-task setting as depicted in Figure \ref{fig:abstract_design}(c). For more details about the SelectiveNet model, readers are advised to read \cite{selective2019}.

\begin{figure*}[t]
\begin{center}
\includegraphics[width=\textwidth, keepaspectratio]{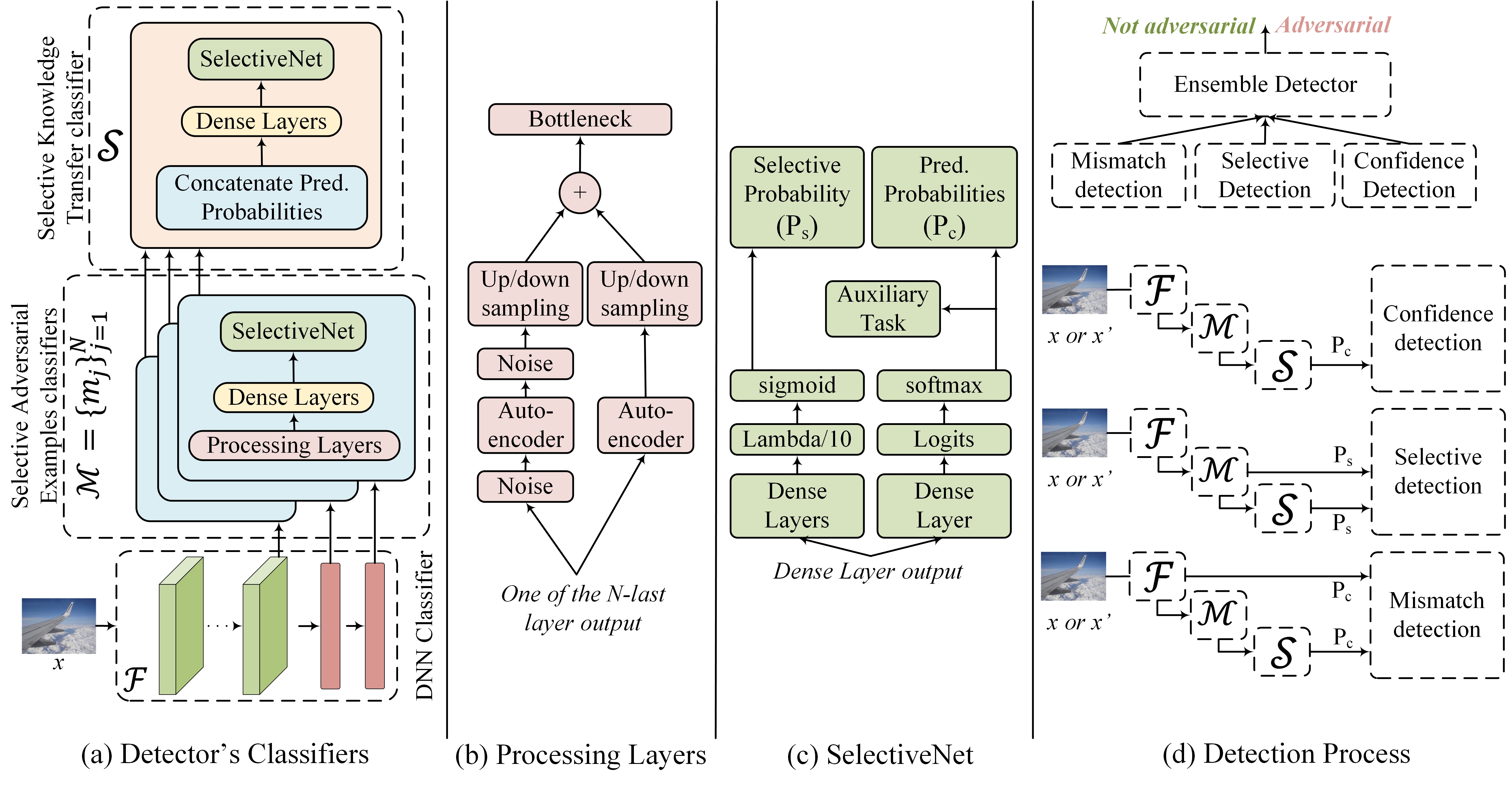}
\end{center}
\caption{(a) Detector architecture. $N$-last representative output of \ac{dnn} is used to build $N$ Selective Adversarial Example classifiers. The confidence output, i.e. pred. probabilities, of the $N$ classifiers is concatenated to be as input for Selective Knowledge Transfer classifier. (b) Feature maps processing blocks. (c) SelectiveNet architecture\cite{selective2019} (d) Detection Process: Selective probabilities ($P_s^{m_j}$ and $P_s^\mathcal{S}$) are used in the Selective detection process. Confidence/Prediction probabilities, ($P_c^{m_j}$ and $P_c^\mathcal{S}$), are used in the confidence detection process. Confidence/Prediction probabilities, ($P_c^\mathcal{F}$ and $P_c^\mathcal{S}$), are used in the mismatch detection process. The total detection is the ensemble of the three detection modules. }
\label{fig:detector_design}
\end{figure*}

\section{\acf{sfad} Method} \label{sec:sfad}

\subsection{Detector design} \label{sec:detector_design}
It is believed that the last $N$ layers in the \ac{dnn} $\mathcal{F}$ have potentials in detecting and rejecting \acp{ae} \cite{melis2017deep,sotgiu2020deep}. In \cite{bendale2016towards} and \cite{melis2017deep}, only the last layer ($N=1$) is utilized to detect \acp{ae}. At this very high level of presentation, \acp{ae} are indistinguishable from samples of the target class. This observation is enhanced when \ac{dnr} \cite{sotgiu2020deep} used the last three layers to build \ac{svm} with \ac{rbf} kernel based classifiers. Unlike other works, in this work, 1) feature maps of the last layers $Z_j$, where $j=\{1,2 \dots,N\}$, are processed. In the aforementioned methods, the representatives of the last layers are not processed and basically the detectors represent another approximation of the baseline classifier which is considered as a weak point. 2) \ac{mtl} is used via the SelctiveNet. \ac{mtl} has an advantage of combining related tasks with one or more loss function(s) and it does better generalization especially with the help of the auxiliary functions. For more details about \ac{mtl}, please refer to these recent review papers \cite{ruder2017overview,vandenhende2020revisiting}. 

In this section, the \acf{sfad} method is demonstrated. As depicted in Figure~\ref{fig:detector_design}(a), \ac{sfad} consists of two main blocks; the selective \acp{ae} classifiers $\mathcal{M}$ block (in blue), where $\mathcal{M}=\{m_j\}_{j=1}^{N}$, and the selective knowledge transfer classifier $\mathcal{S}$ block (in orange). The output of $\mathcal{F}$, $\mathcal{M}$, and $\mathcal{S}$ blocks, i.e. model's uncertainties and confidences, are used in the detection process, as depicted in Figure \ref{fig:abstract_design}(d).

\subsection{Selective \acp{ae} classifiers block} \label{sec:fbm}
As shown in Figure \ref{fig:abstract_design}(a), the aim of $\mathcal{M}$ block is to build $N$ individual classifiers, $\mathcal{M}=\{m_j\}_{j=1}^{N}$. It was shown that perturbation propagation becomes clear when the \ac{dnn} model goes deeper, hence, using $N$-last layers have potential in identifying the \acp{ae}. Unlike works in \cite{melis2017deep,sotgiu2020deep}, we process the representative last $N$-layer(s) outputs $Z_j$ in different ways in order to make clean input features more unique. This will limit the feature space that the adversary uses to craft the \acp{ae} \cite{xu2017feature,meng2017magnet}. Moreover, each of the last $N$-layer output has its own feature space which makes each of the $N$ classifiers $m_j$ to be trained with different feature space. Hence, combining and increasing the number of $N$ will enhance the detection process. 

For simplicity and as recommended in \cite{sotgiu2020deep}, we set $N=3$ in the implemented prototype and hence, each individual layer output is assigned to a classifier as shown in Figure \ref{fig:abstract_design}(a). Let the last $N$ layers' outputs $z_{ji}$ of $x_i$ from $S_k$ are $z_{1i}$, $z_{2i}$, and $z_{Ni}$, respectively, where, $j=\{1,2, \dots, N\}$. $z_{ji}$ are individually the inputs of the $m_j$ classifier. 

The outputs of the $m_j$ classifier are denoted as $m_{ji}(z_{ji})$. Let $\mathcal{Y^\prime}=\mathcal{Y}+1$ be a label space of $m_j$, where the extra label is denoted for the selective status, hence $m_j$ represents a function $m_j:Z_j \rightarrow Y^\prime$ on a distribution $\mathbb{P}(Z_j,Y^\prime)$ over $\mathcal{Z} \times \mathcal{Y^\prime}$. We refer to the selective probability of $m_j$ as $P_s^{m_j}$ and the confidence probabilities of $m_j$ as $P_c^{m_j}$. $m_j$ optimizes the overall loss function
\begin{ceqn}
\begin{equation}
    \mathcal{L}_{m_j} = \alpha \mathcal{L}_{(m_j,g_{m_j})} + (1-\alpha) \mathcal{L}_{h_{m_j}} \text{ , where }\alpha=0.5,
\end{equation}
\end{ceqn}
where $\mathcal{L}_{(m_j,g_{m_j})}$ is the selective loss function of $m_j$, and $\mathcal{L}_{h_{m_j}}$ is the auxiliary loss function of $m_j$ and are calculated as following:
\begin{ceqn}
\begin{equation*}\label{eq:selectiveloss_q}
    \begin{gathered}
        \mathcal{L}_{(m_j,g_{m_j})} \triangleq \hat{r}_{\ell}(m_j,g_{m_j}\mid S_k) + \lambda\, \Psi(c-\hat{\phi}(g_{m_j}\mid S_k)),\\
        \Psi(a) \triangleq \max(0,a)^2,
    \end{gathered}
\end{equation*}
\end{ceqn}

\begin{ceqn}
\begin{equation*}\label{eq:auxiliaryloss_q}
    \mathcal{L}_{h_{m_j}} = \hat{r}(h_{m_j}\mid S_k) = \frac{1}{k}\sum_{i=1}^{k}\ell(h_m(z_{ji}),y_i).
\end{equation*}
\end{ceqn}

Studying the value of $\alpha$ is out of the paper scope, but other task balancing methods, may be applied like,  uncertainty \cite{kendall2018multi}, GradNorm \cite{chen2017gradnorm}, DWA \cite{liu2019end}, DTP \cite{guo2018dynamic}, and MGDA \cite{sener2018multi}.

\subsection{Selective knowledge transfer block} \label{sec:ktm}
The block $\mathcal{S}$ aims at building selective knowledge transfer classifier. It concatenates the confidence values of $Y$ classes of the $\mathcal{M}$ classifiers. The idea behind the block $\mathcal{S}$ is that each set of its input is considered as a special feature of the clean input. Hence, we transfer this knowledge, $m_j$ confidence probabilities, of clean inputs to the classifier. Besides, in the inference time, we believe that \ac{ae} will generate a different distribution of the confidence values and if the \ac{ae} is able to fool one $m_j$, it may not fool the others.

As Figure \ref{fig:detector_design}(a) shows, the confidence probabilities of $m_j$ classifiers are concatenated to be as an input $Q=$ $concat(P_c^{m_1},$ $ P_c^{m_2}, \dots, P_c^{m_N})$ for the selective knowledge transfer block $\mathcal{S}$. The $\mathcal{S}$ classifier consists of one or more dense layer(s) and yields the selective probability of $\mathcal{S}$ as $P_s^\mathcal{S}$ and the confidence probabilities of $\mathcal{S}$ as $P_c^\mathcal{S}$. $\mathcal{S}$ represents a function $\mathcal{S}:Q \rightarrow Y^\prime$ on a distribution $\mathbb{P}(Q,Y^\prime)$ over $\mathcal{Q} \times \mathcal{Y^\prime}$. Hence, it optimizes the following loss function

\begin{ceqn}
\begin{equation}
    \mathcal{L}_\mathcal{S} = \alpha \mathcal{L}_{(\mathcal{S},g_\mathcal{S})} + (1-\alpha) \mathcal{L}_{h_\mathcal{S}} \text{ , where }\alpha=0.5,
\end{equation}
\end{ceqn}

where $\mathcal{L}_{(\mathcal{S},g_\mathcal{S})}$ is the selective loss function of $\mathcal{S}$, and $\mathcal{L}_{h_\mathcal{S}}$ is the auxiliary loss function of $\mathcal{S}$ and are calculated as following:

\begin{ceqn}
\begin{align*}\label{eq:selectiveloss_t}
    &\begin{gathered}
        \mathcal{L}_{(\mathcal{S},g_\mathcal{S})} \triangleq \hat{r}_{\ell}(\mathcal{S},g_\mathcal{S}\mid S_k) + \lambda \Psi(c-\hat{\phi}(g_\mathcal{S}\mid S_k)),\\
        \Psi(a) \triangleq \max(0,a)^2.
    \end{gathered}
\end{align*}
\end{ceqn}

\begin{ceqn}
\begin{equation*}\label{eq:auxiliaryloss_t}
    \mathcal{L}_{h_\mathcal{S}}  = \hat{r}(h_\mathcal{S}\mid S_k) = \frac{1}{k}\sum_{i=1}^{k}\ell(h_\mathcal{S}(q_i),y_i).
\end{equation*}
\end{ceqn}

\subsection{Feature maps processing} \label{sec:feature_processing_blocks}
As depicted in Figure \ref{fig:detector_design}(b) each selective classifier consists of different processing blocks; auto-encoder block, up/down-sampling block, bottleneck block, and noise block. These blocks aim at giving distinguishable features for input samples to let the detector recognize the \acp{ae} efficiently. 

\textbf{Auto-encoder.} Auto-encoders are widely used as a reconstruction tool and its loss is used as a score for different tasks. For instance, it is used in the detection process of \acp{ae} in \cite{meng2017magnet}. It is believed that \acp{ae} gave higher reconstruction loss than clear images. This process is a.k.a attention mechanism \cite{zhang2019scan,zhang2020auxiliary} and it is used to focus on better representation of input features especially on the shallow classifiers. 

\textbf{Up/down-sampling.} Up sampling and down sampling are used in different deep classifiers \cite{wang2017residual,liu2019end}. The aim of down sampling, a.k.a pooling layers in \ac{nn}, is to gather the global information of the input signal. Hence, if we consider the clean input signal as a signal that has global information and then we expand the global information by bi-linear up sampling and then down sample by average pooling, we will measure the ability of global information reconstruction of the input signal. Besides, this process can be seen as a use case of the reconstruction process. 

\textbf{Noise.} Adding noise has a potential impact in making \ac{nn} more robust against \acp{ae} and it has been used in many defense methods \cite{lecuyer2019certified,liu2018towards,liu2020does}. In this work, we add a branch in the classifier that adds small Gaussian noise to the input signal before and after the auto-encoder block. Then, the noised and clean input features are concatenated before the bottleneck block. 

\textbf{Bottleneck.} The bottleneck block \cite{he2016deep} consists of three convolutional layers; 1$\times$1, 3$\times$3, and 1$\times$1 convolutional layers. The bottleneck name came from the fact that the 3$\times$3 convolutional layer is left as a bottleneck between 1$\times$1 convolutional layers. It is mainly designed for efficiency purposes but according to \cite{zhang2019scan,zhang2019your} it is very effective in building shallow classifiers which helps having better representation of input signal.

\subsection{Detection process}
As depicted in Figure \ref{fig:detector_design}(d), the output of baseline model $P_c^\mathcal{F}$, the outputs of $\mathcal{M}$ block, $P_s^{m_j}$ and $P_c^{m_j}$, and the output of $\mathcal{S}$ block, $P_s^\mathcal{S}$ and $P_c^\mathcal{S}$, are used in the ensemble detection process.

\begin{enumerate}
\item \textit{Thresholds: } after finishing the training of the detectors' classifiers, the following thresholds are identified: 
\begin{itemize}
    \item confidence threshold $th_c^{m_j}$ for each selective \acp{ae} classifier $m_j$.
    \item confidence threshold $th_c^{\mathcal{S}}$ for the $\mathcal{S}$ classifier.
    \item the final confidence threshold value 
    
    \begin{ceqn}
        \begin{equation*}
            th_c=\max (th_c^\mathcal{S}, th_c^{m_1}, th_c^{m_2}, ..., th_c^{m_N}) 
        \end{equation*}
    \end{ceqn}
    \item selective threshold $th_s^{m_j}$ for each selective \acp{ae} classifier $m_j$.
    \item selective threshold $th_s^\mathcal{S}$ for the $\mathcal{S}$ classifier.    
\end{itemize}
Following the steps in \cite{sotgiu2020deep}, we select our thresholds at a level when 10\% (at most) of clean inputs can be rejected by the ensemble detection.
\item \textit{Confidence detection:} is set to $1$ if $max(P_c^\mathcal{S})<th_c$ and is set to $0$ otherwise, where 1 means adversarial input.
\item \textit{Selective detection:} is set to $1$ if $P_s^\mathcal{S}<th_s^\mathcal{S}$ or $P_s^{m_1} < th_s^{m_1}$ or \dots \text{ or } $P_s^{m_N} < th_s^{m_N}$ and is set to $0$ otherwise.
\item \textit{Mismatch detection:} is set to $1$ if $\argmax $ $(P_c^\mathcal{S}) \neq \argmax $ $(P_c^\mathcal{F})$ and is set to $0$ otherwise.
\item \textit{Ensemble detection:} The input sample is \textit{adversarial} if it is detected in confidence, selective, or mismatch detection process.
\end{enumerate}

\section{Experimental settings}
\subsection{Datasets}
The proposed prototype is evaluated on CNN models trained with two popular datasets; MNIST \cite{lecun1998gradient}  and \cite{krizhevsky2009learning} CIFAR10. 

MNIST is hand-written digit recognition dataset with 70000 images (60000 for training and 10000 for testing) and ten classes and CIFAR10 is an object recognition dataset with 60000 images (50000 for training and 10000 for testing) ten classes. 

\subsection{Baseline classifiers}
For the baseline models, two CNN models are trained; one for MNIST and one for CIFAR10. For MNIST, we trained 6-layer CNN with 98.73\% accuracy while for CIFAR10 we trained 8-layer CNN with 89.11\% accuracy. The classifier's architectures for MNIST and CIFAR10 are shown in Table \ref{tab:mnist_arch} and Table \ref{tab:cifar10_arch}, respectively.

\begin{table}[!htb]
\centering
\caption{MNIST baseline classifier architecture}
\label{tab:mnist_arch}
\begin{tabular}{m{0.60\columnwidth}|M{0.30\columnwidth}}
\hline
     Layer & Description \\ \hline \hline
     Conv2D + ReLU & 32 filters ($3\times3$) \\ \hline
     Conv2D + ReLU + Max Pooling($2\times2$) & 32 filters ($3\times3$) \\ \hline
     Conv2D + ReLU & 64 filters ($3\times3$) \\ \hline
     Conv2D + ReLU + Max Pooling($2\times2$) & 64 filters ($3\times3$) \\ \hline 
     Dense + ReLU + Dropout ($p=0.3$) & 256 units \\ \hline
     Dense + ReLU & 256 units \\ \hline
     Softmax & 10 classes \\ \hline
\end{tabular}
\end{table}

\begin{table}[!htb]
\centering
\caption{CIFAR10 baseline classifier architecture}
\label{tab:cifar10_arch}
\begin{tabular}{m{0.60\columnwidth}|M{0.30\columnwidth}}
\hline
     Layer & Description \\ \hline \hline
     Conv2D + BatchNorm + ReLU & 64 filters ($3\times3$) \\ \hline
     Conv2D + BatchNorm + ReLU + Max Pooling($2\times2$) + Dropout ($p=0.1$) & 64 filters ($3\times3$) \\ \hline
     Conv2D + BatchNorm + ReLU & 128 filters ($3\times3$) \\ \hline
     Conv2D + BatchNorm + ReLU + Max Pooling($2\times2$) + Dropout ($p=0.2$) & 128 filters ($3\times3$) \\ \hline 
     Conv2D + BatchNorm + ReLU & 256 filters ($3\times3$) \\ \hline
     Conv2D + BatchNorm + ReLU + Max Pooling($2\times2$) + Dropout ($p=0.3$) & 256 filters ($3\times3$) \\ \hline
     Conv2D + BatchNorm + ReLU + Max Pooling($2\times2$) + Dropout ($p=0.4$) & 512 filters ($3\times3$) \\ \hline
     Dense & 512 units \\ \hline
     Softmax & 10 classes \\ \hline
\end{tabular}
\end{table}

In order to evaluate the proposed prototypes against gray-box attacks, we consider that the adversaries know the training dataset and the model outputs and do not know the baseline model architectures. Hence, Table \ref{tab:mnist_arch_gray} and Table \ref{tab:cifar10_arch_gray} show the two alternative architectures for MNIST and CIFAR10 classifiers. For MNIST, the classification accuracies are 98.37\% and 98.69\% for Model \#2 and Model \#3, respectively. While for CIFAR10, the classification accuracies are 86.93\% and 88.38\% for Model \#2 and Model \#3 respectively.

\begin{table}[!htb]
\centering
\caption{MNIST classifiers architectures for gray-box setting}
\label{tab:mnist_arch_gray}
\begin{tabular}{c| m{0.50\columnwidth}|M{0.25\columnwidth}}
\hline
     Model & Layer & Description \\ \hline \hline
     \multirow{7}{*}{\rotatebox[origin=c]{90}{Model \#2}} & Conv2D + BatchNorm + ReLU & 64 filters ($3\times3$) \\ \cline{2-3}
     & Conv2D + BatchNorm + ReLU + Max Pooling($2\times2$) + Dropout ($p=0.5$) & 64 filters ($3\times3$) \\ \cline{2-3}
     & Dense + BatchNorm + ReLU + Dropout ($p=0.5$) & 128 units \\ \cline{2-3}
     & Softmax & 10 classes \\ \hline \hline
     \multirow{8}{*}{\rotatebox[origin=c]{90}{Model \#3}} & Conv2D + ReLU + Max Pooling($2\times2$) & 32 filters ($3\times3$) \\ \cline{2-3}
     & Conv2D + ReLU + Max Pooling($2\times2$) & 64 filters ($3\times3$) \\ \cline{2-3}
     & Dense + ReLU + Dropout ($p=0.5$) & 256 units \\ \cline{2-3}
     & Dense + ReLU & 256 units \\ \cline{2-3}
     & Softmax & 10 classes \\ \hline
\end{tabular}
\end{table}
\begin{table}[!htb]
\centering
\caption{CIFAR10 classifiers architectures for gray-box setting}
\label{tab:cifar10_arch_gray}
\begin{tabular}{c | m{0.50\columnwidth}|M{0.25\columnwidth}}
\hline
     Model & Layer & Description \\ \hline \hline
     \multirow{13}{*}{\rotatebox{90}{Model \#2}} & Conv2D + BatchNorm + ReLU & 32 filters ($3\times3$) \\ \cline{2-3}
     & Conv2D + BatchNorm + ReLU + Max Pooling($2\times2$) & 32 filters ($3\times3$) \\ \cline{2-3}
     & Conv2D + BatchNorm + ReLU & 64 filters ($3\times3$) \\ \cline{2-3}
     & Conv2D + BatchNorm + ReLU + Max Pooling($2\times2$)& 64 filters ($3\times3$) \\ \cline{2-3}
     & Conv2D + BatchNorm + ReLU & 128 filters ($3\times3$) \\ \cline{2-3}
     & Conv2D + BatchNorm + ReLU + Max Pooling($2\times2$)  + Dropout ($p=0.4$) & 128 filters ($3\times3$) \\ \cline{2-3}
     & Dense + BatchNorm + ReLU + Dropout ($p=0.5$) & 512 units \\ \cline{2-3}
     & Softmax & 10 classes \\ \hline \hline
     \multirow{10}{*}{\rotatebox[origin=c]{90}{Model \#3}} & Conv2D + ReLU & 64 filters ($3\times3$) \\ \cline{2-3}
     & Conv2D + ReLU + Max Pooling($2\times2$)& 64 filters ($3\times3$) \\ \cline{2-3}
     & Conv2D + ReLU & 128 filters ($3\times3$) \\ \cline{2-3}
     & Conv2D + ReLU + Max Pooling($2\times2$) & 128 filters ($3\times3$) \\ \cline{2-3}
     & Dense + ReLU  + Dropout ($p=0.5$) & 256 filters ($3\times3$) \\ \cline{2-3}
     & Dense + ReLU & 256 filters ($3\times3$) \\ \cline{2-3}
     & Softmax & 10 classes \\ \hline
\end{tabular}
\end{table}
\subsection{\ac{sfad} Settings}
As described in Section \ref{sec:sfad} and Figure \ref{fig:detector_design}, we introduce here the implementation details for the detector components.
\subsubsection{Selective \acp{ae} classifiers block}
\begin{figure}[!h]\vspace{-7mm}
\begin{center}
\includegraphics[width=\linewidth, keepaspectratio]{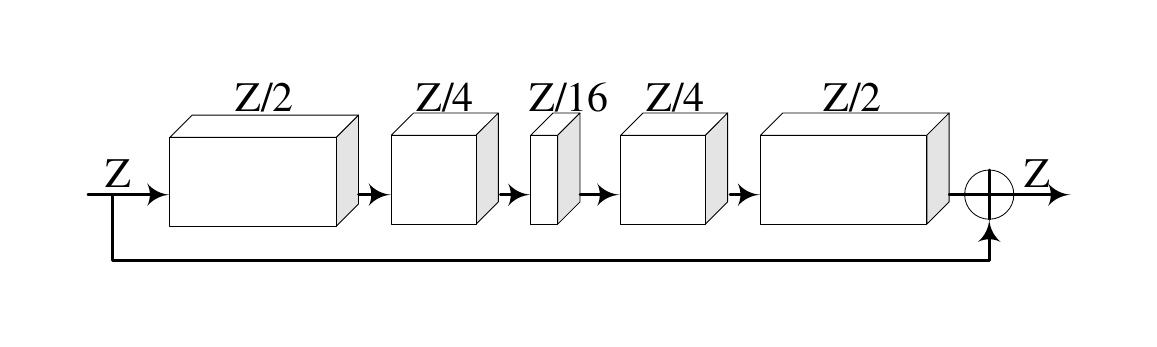}\vspace{-7mm}
\end{center}
\caption{Autoencoder architecture}
\vspace{-2mm}
\label{fig:autoencoder}
\end{figure}

It consists of an autoencoder, up/down sampling, bottleneck, and noise layers. Each has the following architecture:\\
\textbf{Autoencoder.} As shown in Figure \ref{fig:autoencoder}, let the input size be $Z \times w \times h$. In the encoding process, the number of $3\times3$-kernel filters are set to $Z/2$, $Z/4$, and $Z/16$, respectively. In the decoding process, the number of filters $Z$ are symmetrically restored. Finally, to maintain the input samples characteristics that we have before autoencoding, the input is added/summed to the output of the autoencoder.

\begin{figure}[!h]\vspace{-10mm}
\begin{center}
\includegraphics[width=\linewidth, keepaspectratio]{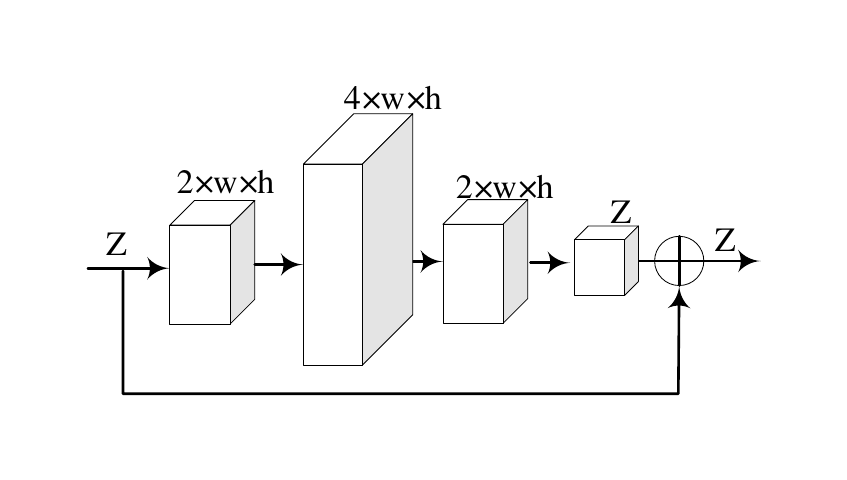}\vspace{-7mm}
\end{center}
\caption{Up/down-sampling architecture}
\vspace{-2mm}
\label{fig:upsampling}
\end{figure}

\textbf{Up/down-sampling.} As shown in Figure \ref{fig:upsampling}, let the input size be $Z \times w \times h$. The input size is doubled by bilinear up sampling in the first two consecutive layers and then restored by average pooling in the last two layers. Finally, to maintain the features before up/down sampling, the input is added to the output of up/down-sampling.

\begin{figure}[!h]
\begin{center}
\includegraphics[width=\linewidth, keepaspectratio]{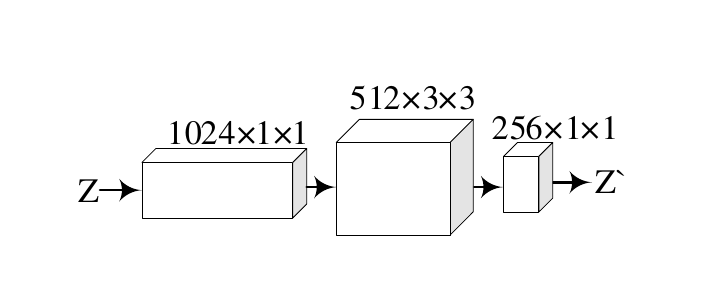}\vspace{-7mm}
\end{center}
\caption{Bottleneck architecture}
\label{fig:bottleneck}
\end{figure}

\textbf{Bottleneck.} It is a three-convolutional layer module with kernels of size $1 \times 1$, $3 \times 3$, and $1\times 1$. The architecture of the bottleneck layers are shown in Figure \ref{fig:bottleneck}. The number of the filters for each layer is 1024, 512, and 256.

\textbf{Noise.} For this layer, the GaussianNoise layer model from Keras library is used with small standard variation of 0.05.

\textbf{Dense layers.} A dense layer with 512 output is used followed by batch normalization and ReLU activation function. 

\textbf{SelectiveNet.} A dense layer with 512 outputs is used followed by batch normalization and ReLU activation function. After that, as original SelectiveNet's implementation suggests, a layer that divides the result of the previous layer by 10 is used as a normalization step. Finally, a dense layer of one output is used with sigmoid activation function. We set $\lambda=32$, $c=1$ for MNIST and $c=0.9$ for CIFAR10, and coverage threshold to 0.995 for MNIST and 0.9 for CIFAR10. More details about selectiveNet hyper-parameters are found in \cite{selective2019}.

\subsubsection{Selective Knowledge Transfer block.} It consists of one dense layer with 128 outputs followed by batch normalization and ReLU activation function. The selective task of the knowledge transfer block consists of a dense layer with 128 outputs followed by batch normalization and ReLU activation function. After that a normalisation layer that divides the result of the previous layer by 10 is used as recommended by the original implementation of SelectiveNet. Finally, a dense layer of one output is used with sigmoid activation function. We set $\lambda=32$, $c=1$ for MNIST and $c=0.9$ for CIFAR10, and coverage threshold to 0.7 for MNIST and CIFAR10. More details about selectiveNet hyper-parameters are found in \cite{selective2019}.

\subsection{Threat Model, attacks, and state-of-the-art detectors} 
\textbf{Threat Model}. We follow one of the threat models presented in \cite{biggio2013evasion,carlini2017adversarial}; Zero-Knowledge adversary threat model. It is assumed that the adversary has no knowledge that a detector is deployed and he generates the white-box attacks with the knowledge of the baseline classifier. For cases when an adversary has perfect or limited knowledge of the detector, we assume that the adversary's work will be so hard since \ac{sfad} adopts ensemble detection, and hence, we leave this as future work. Instead, we tested \ac{sfad} robustness with the recommended high confidence strong attack, a variant of \ac{cw} attack, that is rarely tested in other detectors.

\textbf{Attacks}. We tested the proposed model with different types of attacks. For white-box attacks; $L_\infty$-norm attacks are used; \acs{fgsm} \cite{goodfellow2014explaining}, \acs{pgd} \cite{madry2017towards}, and \ac{cw} \cite{carlini2017towards} attacks. Besides, $L_2$-norm \ac{df} \cite{moosavi2016deepfool} attack is used as well in the testing process. For \acs{fgsm} and \acs{pgd} attacks, the epsilon is set to different values from 0.03 to 0.4. For the \ac{cw} attack, the number of iterations is set to be 1000 (strong case). 
Other attacks' hyper-parameters are set to default as defined in the ART \cite{art2018} library. The \acl{hca} \cite{carlini2017adversarial} is also used to test the robustness of the proposed model. For the black-box attacks, \ac{ta} \cite{kotyan2019adversarial}, \ac{pa} \cite{su2019one}, and \ac{st} \cite{engstrom2019exploring} attacks are used in the testing process. The threshold parameter for \ac{ta} and \ac{pa} is set to 10 for MNIST and to ART defaults for CIFAR. The translation and rotation values of \ac{st} attack are set to 10 and 60 for MNIST and to 8 and 30 for CIFAR, respectively. For the comparison with the state of the art algorithms, more black box attacks are considered like \ac{sa} \cite{andriushchenko2020square}, and \hop attack (HSJA) \cite{chen2020hopskipjumpattack}. For \ac{sa}, the epsilon $\epsilon$ is set to 0.4 to 0.1 for MNIST and CIFAR10 respectively. For \hop attack, untargeted and unmasked attack is considered, besides, 40 and 100 were set for iteration steps and maximum evaluations, respectively.

\textbf{Comparison with existing detectors}.
Supervised and unsupervised detectors are used to compare \ac{sfad} with state-of-the-art techniques. Supervised methods like \ac{kd}+\ac{bu} \cite{feinman2017detecting}, \ac{lid} \cite{ma2018characterizing}, and RAID \cite{eniser2020raid} are compared with \ac{sfad}. While unsupervised methods like FS \cite{xu2017feature}, MagNet \cite{meng2017magnet}, \ac{nic} \cite{ma2019nic}, and \ac{dnr} \cite{sotgiu2020deep} are also considered in the comparisons.

\begin{table*}[t]
\centering
\caption{Performance accuracy (\%) of different detection processes against white-box attacks($\epsilon$) on MNIST and CIFAR10 datasets at FP=10\%. Prediction row is related to baseline \ac{dnn} top-1 accuracy of not detected \acp{ae}. Total = Ensemble Detection + Prediction. }
\label{tab:results_white_mnist}
\resizebox{\textwidth}{!}{%
\begin{tabular}{|c|p{0.08\textwidth}|M{0.045\textwidth}|M{0.045\textwidth}|M{0.045\textwidth}|M{0.045\textwidth}|M{0.045\textwidth}|M{0.05\textwidth}|M{0.045\textwidth}|M{0.045\textwidth}|M{0.045\textwidth}|M{0.045\textwidth}|M{0.045\textwidth}|M{0.045\textwidth}|M{0.045\textwidth}|M{0.045\textwidth}|M{0.045\textwidth}|}
\cline{3-17}
\multicolumn{2}{c|}{}                                                                                                                   & \begin{tabular}[c]{@{}c@{}}FGSM\\(0.05)\end{tabular} & \begin{tabular}[c]{@{}c@{}}\acs{fgsm}\\(0.075)\end{tabular} & \begin{tabular}[c]{@{}c@{}}\acs{fgsm}\\(0.1)\end{tabular} & \begin{tabular}[c]{@{}c@{}}\acs{fgsm}\\(0.2)\end{tabular} & \begin{tabular}[c]{@{}c@{}}\acs{fgsm}\\(0.4)\end{tabular} & \begin{tabular}[c]{@{}c@{}}\acs{fgsm}\\(AVG)\end{tabular} & \begin{tabular}[c]{@{}c@{}}\acs{pgd}\\(0.05)\end{tabular} & \begin{tabular}[c]{@{}c@{}}\acs{pgd}\\(0.075)\end{tabular} & \begin{tabular}[c]{@{}c@{}}\acs{pgd}\\(0.1)\end{tabular} & \begin{tabular}[c]{@{}c@{}}\acs{pgd}\\(0.2)\end{tabular} & \begin{tabular}[c]{@{}c@{}}\acs{pgd}\\(0.4)\end{tabular} & \begin{tabular}[c]{@{}c@{}}\acs{pgd}\\(AVG)\end{tabular} & \acs{df}                            & CW                                                      \\ \hline
\multicolumn{1}{|c|}{}                        & \begin{tabular}[c]{@{}l@{}}Baseline\\\ac{dnn}\end{tabular}                                  & 96.31                                                  & 92.93                                                   & 87.2                                                  & 28.04                                                 & 7.91                                                  & -                                                     & 95.18                                                 & 85.84                                                  & 56.91                                                & 0                                                    & 0                                                    & -                                                    & 4.68                          & 38.97                                                  \\ \hhline{|~================|} 
\multicolumn{1}{|c|}{}                        & \begin{tabular}[c]{@{}l@{}}Selective\\Detection\end{tabular} & 18.82                          & 25.02                           & 33.08                         & 82.29                         & 98.4                          & 51.52                         & 20.49                         & 32.91                          & 51.81                        & 58.07                        & 48.17                        & 42.29                        & 95.7  & 43.71  \\ \cline{2-17} 
\multicolumn{1}{|c|}{}                        & \begin{tabular}[c]{@{}l@{}}Confidence\\Detection\end{tabular}                        & 8.66                                                   & 13.74                                                   & 20.75                                                 & 74.18                                                 & 98.66                                                 & 43.2                                                  & 10.31                                                 & 21                                                     & 42.73                                                & 54.14                                                & 47.41                                                & 35.12                                                & 94.96                         & 42.78                                                  \\ \cline{2-17} 
\multicolumn{1}{|c|}{}                        & \begin{tabular}[c]{@{}l@{}}Mismatch\\Detection\end{tabular}  & 3.07                           & 5.66                            & 9.77                          & 43.43                         & 63.93                         & 25.17                         & 3.85                          & 10.5                           & 28.61                        & 26.86                        & 18.69                        & 17.7                        & 59.47 & 44.99  \\ \cline{2-17}
\multicolumn{1}{|c|}{}                        &
\cellcolor[HTML]{DEDEDE}\begin{tabular}[c]{@{}l@{}}Ensemble\\Detection\end{tabular}  & \cellcolor[HTML]{DEDEDE}22.97                           & \cellcolor[HTML]{DEDEDE}31.24                            & \cellcolor[HTML]{DEDEDE}40.88                          & \cellcolor[HTML]{DEDEDE}88.83                         & \cellcolor[HTML]{DEDEDE}99.8                         & \cellcolor[HTML]{DEDEDE}56.74                         & \cellcolor[HTML]{DEDEDE}25.6                          & \cellcolor[HTML]{DEDEDE}41.91                           & \cellcolor[HTML]{DEDEDE}66.71                        & \cellcolor[HTML]{DEDEDE}68.09                        & \cellcolor[HTML]{DEDEDE}58.93                        & \cellcolor[HTML]{DEDEDE}52.25                        & \cellcolor[HTML]{DEDEDE}99.14                      & \cellcolor[HTML]{DEDEDE}61.21  \\ \cline{2-17}
\multicolumn{1}{|c|}{}                        & Prediction                                                                             & 76.99                                                  & 68.64                                                   & 58.74                                                 & 9.03                                                  & 0                                                     & 42.68                                                 & 74.34                                                 & 57.72                                                  & 31.69                                                & 0                                                    & 0                                                    & 32.75                                               & 0.19                          & 37.44                                               \\ \cline{2-17} 
\multicolumn{1}{|c|}{\multirow{-10}{*}{\rotatebox[origin=c]{90}{MNIST}}} & \cellcolor[HTML]{DEDEDE}Total                                                          & \cellcolor[HTML]{DEDEDE}99.96                          & \cellcolor[HTML]{DEDEDE}99.88                           & \cellcolor[HTML]{DEDEDE}99.62                         & \cellcolor[HTML]{DEDEDE}97.86                         & \cellcolor[HTML]{DEDEDE}99.8                          & \cellcolor[HTML]{DEDEDE}99.42                         & \cellcolor[HTML]{DEDEDE}99.94                         & \cellcolor[HTML]{DEDEDE}99.63                          & \cellcolor[HTML]{DEDEDE}98.4                         & \cellcolor[HTML]{DEDEDE}68.09                        & \cellcolor[HTML]{DEDEDE}58.93                        & \cellcolor[HTML]{DEDEDE}85                           & \cellcolor[HTML]{DEDEDE}99.33 & \cellcolor[HTML]{DEDEDE}98.65  \\ \hline
\multicolumn{1}{|c|}{}                        & \begin{tabular}[c]{@{}l@{}}Baseline\\\ac{dnn}\end{tabular}                                  & 14.09                                                  & 13.44                                                   & 12.25                                                 & 10.5                                                  & 9.75                                                  & -                                                    & 0.43                                                  & 0.28                                                   & 0.22                                                 & 0.16                                                 & 0.17                                                 & -                                                    & 4.79                          & 20.95                                             \\ \hhline{|~================|} 
\multicolumn{1}{|c|}{}                        & \begin{tabular}[c]{@{}l@{}}Selective \\ Detection\end{tabular} & 41.31                          & 46.8                            & 53.78                         & 71.34                         & 9.83                          & 44.61                         & 33.6                          & 33.18                          & 30.52                        & 22.66                        & 18.65                        & 27.72                        & 39.06 & 37.41 \\ \cline{2-17} 
\multicolumn{1}{|c|}{}                        & \begin{tabular}[c]{@{}l@{}}Confidence \\ Detection\end{tabular}                        & 69.3                                                   & 80.61                                                   & 87.35                                                 & 99.09                                                 & 99.99                                                 & 87.27                                                 & 43.33                                                 & 54.69                                                  & 60.86                                                & 66.83                                                & 67.68                                                & 58.68                                                & 85.76                         & 65.47                                              \\ \cline{2-17} 
\multicolumn{1}{|c|}{}                        & \begin{tabular}[c]{@{}l@{}}Mismatch \\ Detection\end{tabular}  & 25.49                          & 35.15                           & 42.88                         & 44.41                         & 59.66                         & 41.52                         & 0                             & 0                              & 0.02                         & 0.11                         & 1.32                         & 0.29                         & 37.5  & 34.42\\ \cline{2-17}
\multicolumn{1}{|c|}{}                        &
\cellcolor[HTML]{DEDEDE}\begin{tabular}[c]{@{}l@{}}Ensemble\\Detection\end{tabular}  & \cellcolor[HTML]{DEDEDE}72.07                           & \cellcolor[HTML]{DEDEDE}81.84                            & \cellcolor[HTML]{DEDEDE}88.42                          & \cellcolor[HTML]{DEDEDE}99.41                         & \cellcolor[HTML]{DEDEDE}100                         & \cellcolor[HTML]{DEDEDE}88.34                         & \cellcolor[HTML]{DEDEDE}57.57                          & \cellcolor[HTML]{DEDEDE}63.59                           & \cellcolor[HTML]{DEDEDE}66.67                        & \cellcolor[HTML]{DEDEDE}68.77                        & \cellcolor[HTML]{DEDEDE}68.74                        & \cellcolor[HTML]{DEDEDE}65.07                        & \cellcolor[HTML]{DEDEDE}88.45                      & \cellcolor[HTML]{DEDEDE}69.93  \\ \cline{2-17}
\multicolumn{1}{|c|}{}                        & Prediction                                                                             & 6.94                                                   & 3.28                                                    & 1.39                                                  & 0.02                                                  & 0                                                     & 2.33                                                  & 0.34                                                  & 0.13                                                   & 0.07                                                 & 0.06                                                 & 0.04                                                 & 0.13                                                 & 1.35                          & 20.09                                         \\ \cline{2-17} 
\multicolumn{1}{|c|}{\multirow{-10}{*}{\rotatebox[origin=c]{90}{CIFAR}}} & \cellcolor[HTML]{DEDEDE}Total                                                          & \cellcolor[HTML]{DEDEDE}79.01                          & \cellcolor[HTML]{DEDEDE}85.12                           & \cellcolor[HTML]{DEDEDE}89.81                         & \cellcolor[HTML]{DEDEDE}99.43                         & \cellcolor[HTML]{DEDEDE}100                           & \cellcolor[HTML]{DEDEDE}90.67                         & \cellcolor[HTML]{DEDEDE}57.91                         & \cellcolor[HTML]{DEDEDE}63.72                          & \cellcolor[HTML]{DEDEDE}66.74                        & \cellcolor[HTML]{DEDEDE}68.83                        & \cellcolor[HTML]{DEDEDE}68.78                        & \cellcolor[HTML]{DEDEDE}65.2                         & \cellcolor[HTML]{DEDEDE}89.8  & \cellcolor[HTML]{DEDEDE}90.02 \\ \hline
\end{tabular}
}
\end{table*}

\section{Experimental Results and Discussion}
In this section, we evaluate the prototype of SFAD against different types of attack scenarios and datasets and then, we provide a comparison discussion with state-of-the-art detectors. Performance results on successful attacks only are also discussed. Besides, the proposed approach is tested with different $N$ settings and the results are discussed. In order to emphasize the advantages of SFAD's feature processing components, we provide an ablation study for each component. Finally, performance results on different rejection rates, i.e. false positive rates are shown. As a reminder, we use only the last three representative layers ($N=3$) to build three selective \acp{ae} classifiers since the aim is to prove the concept of the approach and if that is changed with the best combination, the detector accuracy will be enhanced accordingly.   

\subsection{Zero-Knowledge (of detectors) adversary white-box attacks}
Table~\ref{tab:results_white_mnist} shows selective, confidence, mismatch, and ensemble detection accuracies of \ac{sfad} prototype for MNIST and CIFAR10 datasets. It also shows the baseline \ac{dnn} prediction accuracy for the \acp{ae} in ``Baseline \ac{dnn}'' row and for the not detected \acp{ae} in ``prediction'' row. The ``Total'' row is the total accuracy of ensemble detection and truly classified/predicted samples. 

For MNIST dataset, the \ac{fgsm} attacks with small epsilon ($\epsilon=0.05,$ $0.075,$ and $0.1$) slightly fooled the baseline classifier and hence their feature space still inside or at the border as of training dataset. The detector shows its ability to reject those samples that are so close to the classes borders and achieves the accuracy of 99.96\%, 99.88\%, and 99.62\% for $\epsilon=(0.05, 0.075, \text{ and}, 0.1)$,  respectively. Similar observation is noticed for \ac{pgd} attacks with small $\epsilon$ values. For larger $\epsilon$ values, \ac{df}, and \ac{cw} attacks, the \acp{ae} are highly able to fool the baseline classifier since adversaries are able to change the MNIST test samples' feature space to lie out of its corresponding class border and hence, for all tested attacks except the \ac{pgd}, the model was able to catch them with accuracy above 98.65\%. While the detector achieves 68.09\% and 58.93\% for \ac{pgd} attacks with $\epsilon=(0.2, \text{ and}, 0.4)$, respectively. Some \ac{pgd} examples' feature space became indistinguishable from the trained samples feature space. That makes \ac{sfad} not able to catch all \acp{ae} and to enhance \ac{sfad}'s performance, the best representative layers combination has to be used as input for the detector. 

For CIFAR10 dataset, \ac{sfad} achieves comparable results with state-of-the-art methods for \ac{fgsm} ($\epsilon=0.1,$ $0.2,$ and $0.4$), \ac{df}, and \ac{cw} attacks. While for \ac{fgsm} ($\epsilon=0.05,$ $\text{and } 0.075$) and \ac{pgd} attacks, the \acp{ae} have, to some extent, indistinguishable feature space than those the detector is trained with. In average, the model achieves accuracy of 65.2\% for \ac{pgd} attacks. 

For both datasets, the effectiveness of selective, confidence, and mismatch detection is obvious. The ability of the two modules to detect the \acp{ae} is increasing when the amount of the perturbations is increasing. When the amount of the perturbations increased in a way that makes the adversarial samples feature space indistinguishable from the training dataset, the ability of these modules to detect the \acp{ae} is decreasing.

\subsection{Black-box attacks}
Table \ref{tab:results_black_mnist_cifar} shows \ac{sfad} prototype's detection accuracy against the \ac{ta} \cite{kotyan2019adversarial}, \ac{pa} \cite{su2019one}, and \ac{st} \cite{engstrom2019exploring} attacks on MNIST and CIFAR10 datasets. The detector is able to catch the \acp{ae} with very high accuracy, higher than 97.56\% and 93.97\% for MNIST and  CIFAR10, respectively. It is clear that the selective, confidence, and mismatch modules complement each other. The black-box attacks significantly change the samples features that facilitate the confidence module detection process. While the ability of selective module is limited for \ac{ta} and \ac{pa} attacks since these attacks change one or more pixels within a threshold that is in a variation of the input sample and yield \acp{ae} that are so close to clean samples. 

\begin{table*}
\centering
\caption{Performance accuracy (\%) of different detection processes against black-box attacks on MNIST and CIFAR10 datasets at FP=10\%. Prediction column is related to baseline \ac{dnn} top-1 accuracy of not detected \acp{ae}. Total = Ensemble Detection + Prediction. }
\label{tab:results_black_mnist_cifar}
\resizebox{0.9\textwidth}{!}{%
\begin{tabular}{|M{0.06\textwidth}|l|c||M{0.08\textwidth}|M{0.09\textwidth}|M{0.08\textwidth}|M{0.08\textwidth}|M{0.08\textwidth}|M{0.07\textwidth}|} 
\hline
\multirow{3}{*}{Dataset} & \multirow{3}{*}{Attack} & \multirow{3}{*}{Baseline \ac{dnn}} & \multicolumn{6}{c|}{Proposed}                                                                                                  \\ 
\cline{4-9}
                         &                         &                                      & Selective Detection & Confidence Detection  & Mismatch Detection & \cellcolor[HTML]{DEDEDE}Ensemble Prediction & Prediction & \cellcolor[HTML]{DEDEDE}Total  \\ 
\hline
\multirow{3}{*}{MNIST}   & Threshold Attack                      & 77.61                                & 24.36                        & 85.48                         & 42.37           & \cellcolor[HTML]{DEDEDE}85.62            & 14.31           & \cellcolor[HTML]{DEDEDE}99.93           \\ 
\cline{2-9}
                         & Pixel Attack                      & 74.57                                & 24.65                        & 85.57                         & 42.88     & \cellcolor[HTML]{DEDEDE}85.76                   & 14.18           & \cellcolor[HTML]{DEDEDE}99.94           \\ 
\cline{2-9}
                         & Spatial Transformation                      & 22.04                                & 86.7                         & 80.95                         & 34.71          & \cellcolor[HTML]{DEDEDE}94.74              & 2.85            & \cellcolor[HTML]{DEDEDE}97.59           \\ 
\hline \hline
\multirow{3}{*}{CIFAR}   & Threshold Attack                      & 11.29                                & 12.69                        & 92.14                         & 37.11              & \cellcolor[HTML]{DEDEDE}92.62          & 1.35            & \cellcolor[HTML]{DEDEDE}93.97           \\ 
\cline{2-9}
                         & Pixel Attack                      & 11.35                                & 12.48                        & 92.4                          & 37.02     & \cellcolor[HTML]{DEDEDE}92.77                   & 1.39            & \cellcolor[HTML]{DEDEDE}94.16           \\ 
\cline{2-9}
                         & Spatial Transformation                      & 52.58                                & 44.64                        & 68.16                         & 32.44        & \cellcolor[HTML]{DEDEDE}72.54                & 24.03           & \cellcolor[HTML]{DEDEDE}96.57           \\
\hline
\end{tabular}
}%
\end{table*}

\begin{table*}
\centering
\caption{Performance accuracy (\%) of different detection processes against grey-box attacks($\epsilon$) on MNIST dataset at FP=10\%. Prediction column is related to baseline \ac{dnn} top-1 accuracy of not detected \acp{ae}. Total = Ensemble Detection + Prediction. }
\label{tab:mnist_gray_attaks}
\resizebox{\textwidth}{!}{%
\begin{tabular}{|l||M{0.06\textwidth}||M{0.07\textwidth}|M{0.08\textwidth}|M{0.07\textwidth}|M{0.07\textwidth}|M{0.07\textwidth}|M{0.05\textwidth}||M{0.07\textwidth}|M{0.08\textwidth}|M{0.07\textwidth}|M{0.07\textwidth}|M{0.07\textwidth}|M{0.05\textwidth}||} 
\hline
\multirow{2}{*}{Attack ($\epsilon$)} & Model\#1                                  & \multicolumn{6}{c||}{Model\#2}                                                                                            & \multicolumn{6}{c||}{Model\#3}                                                                                             \\ 
\hhline{|~-------------|}
                        & {\cellcolor[rgb]{0.867,0.867,0.867}}Total & Selective Detection & Confidence Detection & Mismatch Detection & Ensemble Detection & Prediction & {\cellcolor[rgb]{0.867,0.867,0.867}}Total & Selective Detection & Confidence Detection & Mismatch Detection & Ensemble Detection & Prediction & {\cellcolor[rgb]{0.867,0.867,0.867}}Total  \\ 
\hline
\acs{fgsm}(0.05)              & {\cellcolor[rgb]{0.867,0.867,0.867}}99.96 & 11.94               & 4.31                 & 0.57               & 13.86 & 86.14      & {\cellcolor[rgb]{0.867,0.867,0.867}}100   & 12.98               & 5.01                 & 0.87          & 15.28      & 84.72      & {\cellcolor[rgb]{0.867,0.867,0.867}}100    \\ 
\hline
\acs{fgsm}(0.075)             & {\cellcolor[rgb]{0.867,0.867,0.867}}99.88 & 13.71               & 5.3                  & 1.08           & 16.23     & 83.77      & {\cellcolor[rgb]{0.867,0.867,0.867}}100   & 15.63               & 6.65                 & 1.68        & 18.84        & 81.14      & {\cellcolor[rgb]{0.867,0.867,0.867}}99.98  \\ 
\hline
\acs{fgsm}(0.1)               & {\cellcolor[rgb]{0.867,0.867,0.867}}99.62 & 16.3                & 6.76                 & 1.66       & 19.59         & 80.41      & {\cellcolor[rgb]{0.867,0.867,0.867}}100   & 19.48               & 9.17                 & 2.62         & 23.66       & 76.29      & {\cellcolor[rgb]{0.867,0.867,0.867}}99.95  \\ 
\hline
\acs{fgsm}(0.2)               & {\cellcolor[rgb]{0.867,0.867,0.867}}97.86 & 46.72               & 30.18                & 12.77     & 51.49          & 48.25      & {\cellcolor[rgb]{0.867,0.867,0.867}}99.74 & 57.56               & 47.39                & 24.75       & 63.32        & 35.59      & {\cellcolor[rgb]{0.867,0.867,0.867}}98.91  \\ 
\hline
\acs{fgsm}(0.4)               & {\cellcolor[rgb]{0.867,0.867,0.867}}99.8  & 96.84               & 96.39                & 58.97        & 99.41       & 0.41       & {\cellcolor[rgb]{0.867,0.867,0.867}}99.82 & 95.88               & 98.81                & 55.88         & 99.89      & 0          & {\cellcolor[rgb]{0.867,0.867,0.867}}99.89  \\ 
\hline
\acs{pgd}(0.05)               & {\cellcolor[rgb]{0.867,0.867,0.867}}99.94 & 11.63               & 4.11                 & 0.53        & 13.57        & 86.43      & {\cellcolor[rgb]{0.867,0.867,0.867}}100   & 13.16               & 5.05                 & 0.95         & 15.49       & 84.49      & {\cellcolor[rgb]{0.867,0.867,0.867}}99.98  \\ 
\hline
\acs{pgd}(0.075)              & {\cellcolor[rgb]{0.867,0.867,0.867}}99.63 & 12.9                & 4.67                 & 0.92       & 15.01        & 84.99      & {\cellcolor[rgb]{0.867,0.867,0.867}}100   & 16.34               & 7                    & 1.74         & 19.77       & 80.21      & {\cellcolor[rgb]{0.867,0.867,0.867}}99.98  \\ 
\hline
\acs{pgd}(0.1)                & {\cellcolor[rgb]{0.867,0.867,0.867}}98.4  & 15.41               & 5.99                 & 1.64        & 18.53        & 81.47      & {\cellcolor[rgb]{0.867,0.867,0.867}}100   & 20.92               & 10.13                & 3.55         & 25.73       & 74.21      & {\cellcolor[rgb]{0.867,0.867,0.867}}99.94  \\ 
\hline
\acs{pgd}(0.2)                & {\cellcolor[rgb]{0.867,0.867,0.867}}68.09 & 47.2                & 32.61                & 15.18         & 54.9      & 44.44      & {\cellcolor[rgb]{0.867,0.867,0.867}}99.34 & 60.85               & 54.55                & 30.25         & 73.26      & 23.85      & {\cellcolor[rgb]{0.867,0.867,0.867}}97.11  \\ 
\hline
\acs{pgd}(0.4)                & {\cellcolor[rgb]{0.867,0.867,0.867}}58.93 & 83.59               & 83.73                & 30.33        & 91.1       & 0.93       & {\cellcolor[rgb]{0.867,0.867,0.867}}92.03 & 77.18               & 57.9                 & 21.84        & 82.78       & 0.2        & {\cellcolor[rgb]{0.867,0.867,0.867}}82.98  \\ 
\hline
\acs{df}                      & {\cellcolor[rgb]{0.867,0.867,0.867}}99.33 & 90.33               & 84.72                & 43.46        & 92.25       & 7.37       & {\cellcolor[rgb]{0.867,0.867,0.867}}99.62 & 91.78               & 91.26                & 57.84        & 96.03       & 2.59       & {\cellcolor[rgb]{0.867,0.867,0.867}}98.62  \\ 
\hline
\acs{cw}                      & {\cellcolor[rgb]{0.867,0.867,0.867}}98.65 & 21.01               & 10.96                & 3.24        & 25.16        & 74.82      & {\cellcolor[rgb]{0.867,0.867,0.867}}99.98 & 31.27               & 19.26                & 6.93         & 38.57       & 61.33      & {\cellcolor[rgb]{0.867,0.867,0.867}}99.9   \\
\hline
\end{tabular}
}\vspace{-1mm}
\end{table*}

\begin{table*}
\centering
\caption{Performance accuracy of different detection processes against grey-box attacks($\epsilon$) on CIFAR10 dataset at FP=10\%. Prediction column is related to baseline \ac{dnn} top-1 accuracy of not detected \acp{ae}. Total = Ensemble Detection + Prediction.  }
\label{tab:cifar_gray_attaks}

\resizebox{\textwidth}{!}{%
\begin{tabular}{|l||M{0.06\textwidth}||M{0.07\textwidth}|M{0.08\textwidth}|M{0.07\textwidth}|M{0.07\textwidth}|M{0.07\textwidth}|M{0.05\textwidth}||M{0.07\textwidth}|M{0.08\textwidth}|M{0.07\textwidth}|M{0.07\textwidth}|M{0.07\textwidth}|M{0.05\textwidth}||} 
\hline
\multirow{2}{*}{Attack ($\epsilon$)} & Model\#1                                  & \multicolumn{6}{c||}{Model\#2}                                                                                            & \multicolumn{6}{c||}{Model\#3}                                                                                             \\ 
\hhline{|~-------------|}
                        & {\cellcolor[rgb]{0.867,0.867,0.867}}Total & Selective Detection & Confidence Detection & Mismatch Detection & Ensemble Detection & Prediction & {\cellcolor[rgb]{0.867,0.867,0.867}}Total & Selective Detection & Confidence Detection & Mismatch Detection & Ensemble Detection & Prediction & {\cellcolor[rgb]{0.867,0.867,0.867}}Total  \\ 
\hline
\acs{fgsm}(0.05)              & {\cellcolor[rgb]{0.867,0.867,0.867}}79.01 & 43.77               & 72.69                & 28.27      & 74.09         & 8.56       & {\cellcolor[rgb]{0.867,0.867,0.867}}82.65 & 48.36               & 73.44                & 27.45       & 76.43        & 7.9        & {\cellcolor[rgb]{0.867,0.867,0.867}}84.33  \\ 
\hline
\acs{fgsm}(0.075)             & {\cellcolor[rgb]{0.867,0.867,0.867}}85.12 & 45.61               & 79                   & 33.18       & 79.68        & 4.26       & {\cellcolor[rgb]{0.867,0.867,0.867}}83.94 & 49.16               & 78.54                & 32.32      & 80.25         & 4.17       & {\cellcolor[rgb]{0.867,0.867,0.867}}84.42  \\ 
\hline
\acs{fgsm}(0.1)               & {\cellcolor[rgb]{0.867,0.867,0.867}}89.81 & 53.11               & 87.11                & 38.82      & 87.48         & 2.03       & {\cellcolor[rgb]{0.867,0.867,0.867}}89.51 & 48.36               & 82.7                 & 32.8       & 83.69         & 2.55       & {\cellcolor[rgb]{0.867,0.867,0.867}}86.24  \\ 
\hline
\acs{fgsm}(0.2)               & {\cellcolor[rgb]{0.867,0.867,0.867}}99.43 & 77.37               & 99.58                & 37.72        & 99.66       & 0.03       & {\cellcolor[rgb]{0.867,0.867,0.867}}99.69 & 60.85               & 96.82                & 26.97      & 96.9         & 0.26       & {\cellcolor[rgb]{0.867,0.867,0.867}}97.16  \\ 
\hline
\acs{fgsm}(0.4)               & {\cellcolor[rgb]{0.867,0.867,0.867}}100   & 5.6                 & 100                  & 62.05       & 100        & 0          & {\cellcolor[rgb]{0.867,0.867,0.867}}100   & 23.04               & 99.99                & 16.8        & 100        & 0          & {\cellcolor[rgb]{0.867,0.867,0.867}}100    \\ 
\hline
\acs{pgd}(0.05)               & {\cellcolor[rgb]{0.867,0.867,0.867}}57.91 & 6.55                & 18.95                & 3.46        & 20.7        & 5.62       & {\cellcolor[rgb]{0.867,0.867,0.867}}26.32 & 7.25                & 16.13                & 4.35        & 18.56        & 4.44       & {\cellcolor[rgb]{0.867,0.867,0.867}}23     \\ 
\hline
\acs{pgd}(0.075)              & {\cellcolor[rgb]{0.867,0.867,0.867}}63.72 & 2.58                & 12.43                & 1.12        & 13.18        & 5.13       & {\cellcolor[rgb]{0.867,0.867,0.867}}18.31 & 3.33                & 8.93                 & 1.96       & 10.36         & 4.31       & {\cellcolor[rgb]{0.867,0.867,0.867}}14.67  \\ 
\hline
\acs{pgd}(0.1)                & {\cellcolor[rgb]{0.867,0.867,0.867}}66.74 & 1.93                & 11.95                & 0.91       & 12.6         & 5.07       & {\cellcolor[rgb]{0.867,0.867,0.867}}17.67 & 2.48                & 8.14                 & 1.96        & 9.6        & 4.3        & {\cellcolor[rgb]{0.867,0.867,0.867}}13.9   \\ 
\hline
\acs{pgd}(0.2)                & {\cellcolor[rgb]{0.867,0.867,0.867}}68.83 & 3.5                 & 18.74                & 2.05         & 19.64       & 4.9        & {\cellcolor[rgb]{0.867,0.867,0.867}}24.54 & 6.13                & 16.45                & 4.05       & 19.35         & 3.92       & {\cellcolor[rgb]{0.867,0.867,0.867}}23.27  \\ 
\hline
\acs{pgd}(0.4)                & {\cellcolor[rgb]{0.867,0.867,0.867}}68.78 & 8.19                & 29.39                & 5.57      & 31.76          & 4.35       & {\cellcolor[rgb]{0.867,0.867,0.867}}36.11 & 23.73               & 48.56                & 13.81        & 54.21       & 2.37       & {\cellcolor[rgb]{0.867,0.867,0.867}}56.58  \\ 
\hline
\acs{df}                      & {\cellcolor[rgb]{0.867,0.867,0.867}}89.8  & 37.27               & 77.65                & 42.68    & 81.2           & 16.17      & {\cellcolor[rgb]{0.867,0.867,0.867}}97.37 & 29.3                & 81.02                & 37.75      & 84.28         & 11.74      & {\cellcolor[rgb]{0.867,0.867,0.867}}96.02  \\ 
\hline
\acs{cw}                      & {\cellcolor[rgb]{0.867,0.867,0.867}}90.02 & 24.86               & 40.69                & 11.3       & 45.35         & 53.29      & {\cellcolor[rgb]{0.867,0.867,0.867}}98.64 & 31.7                & 50.07                & 14.82      & 55.06         & 42.18      & {\cellcolor[rgb]{0.867,0.867,0.867}}97.24  \\
\hline
\end{tabular}
}\vspace{-2mm}
\end{table*}

\subsection{Gray-box attacks} \label{sec:gray-box}
Gray-box scenario assumes that the adversary has only knowledge about the model training data and the output of the \ac{dnn} model. Hence, we trained two models as substitution models named Model\#2 and Model\#3 for MNIST and CIFAR10 as shown in Table \ref{tab:mnist_arch_gray} and Table \ref{tab:cifar10_arch_gray}, respectively. Then, white-box based \acp{ae} are generated using the substitution models. The \ac{sfad} prototype is then tested against these \acp{ae}. For both datasets, it is shown in Table \ref{tab:mnist_gray_attaks} and Table \ref{tab:cifar_gray_attaks} that the perturbations properties generated from one model are transferred to the tested model, Model\#1. 
For MNIST, see Table \ref{tab:mnist_gray_attaks}, \ac{sfad} prediction rate is much better for \ac{pgd} attacks and the prediction rate for other attacks is comparable with the prediction rate of \acp{ae} generated from Model\#1. 
For CIFAR10, see Table \ref{tab:cifar_gray_attaks}, the prediction rate for \ac{cw} and \ac{df} attacks is higher than those attacks that are generated using Model\#1, while the prediction rate for \ac{fgsm} is comparable with the prediction rate for \ac{fgsm} attacks that are generated using Mode1\#1. Unlike other attacks, the \ac{pgd} attacks transferable properties sound to be much stronger and have different feature space, compared to feature space of \acp{ae} that are generated from Model\#1. This reduces the ability of the detector to catch such attacks.

\begin{table*}[t]
\centering
\caption{Detection modules' accuracies (\%) against \textit{successful} white-box and black-box attacks($\epsilon$) on MNIST and CIFAR10 datasets at FP=10\%.}
\label{tab:all_suc_attacks}
\resizebox{\textwidth}{!}{%
\begin{tabular}{|M{0.02\textwidth}|M{0.08\textwidth}||M{0.045\textwidth}|M{0.05\textwidth}|M{0.045\textwidth}|M{0.045\textwidth}|M{0.045\textwidth}|M{0.05\textwidth}|M{0.04\textwidth}|M{0.05\textwidth}|M{0.04\textwidth}|M{0.04\textwidth}|M{0.04\textwidth}|M{0.05\textwidth}|M{0.04\textwidth}|M{0.04\textwidth}||M{0.04\textwidth}|M{0.04\textwidth}|M{0.04\textwidth}|} 
\cline{3-19}
\multicolumn{1}{c}{\multirow{2}{*}{}} & \multirow{2}{*}{}                         & \multicolumn{14}{c||}{White-box attacks}                                                                                                                                   & \multicolumn{3}{c|}{Black-box attacks}                                                                                             \\ 
\cline{3-19}
\multicolumn{1}{c}{}                  &                                           & \acs{fgsm} (0.05)                               & \acs{fgsm} (0.075)                              & \acs{fgsm} (0.1)                                & \acs{fgsm} (0.2)                                & \acs{fgsm} (0.4)                                & \textbf{\acs{fgsm} (AVG)}                                & \acs{pgd} (0.05)                                & \acs{pgd} (0.075)                               & \acs{pgd} (0.1)                                 & \acs{pgd} (0.2)                                 & \acs{pgd} (0.4)                                 & \textbf{\acs{pgd} (AVG)}                                 & \acs{df}                                        & \acs{cw}                                                                               & TA                                        & PA                                        & ST                                         \\ 
\hline
\multirow{4}{*}{\rotatebox[origin=c]{90}{MNIST}}      & Selective Detection                       & 59.89                                     & 60.74                                     & 64.56                                     & 89.4                                      & 98.33                                     & \textbf{74.58}                                     & 57.35                                     & 64.23                                     & 68.05                                     & 58.07                                     & 48.17                                     & \textbf{59.17}                                     & 95.78                                     & 69.16                                                                       & 25.96                                     & 28.04                                     & 87.9                                       \\ 
\cline{2-19}
                                      & Confidence Detection                      & 60.99                                     & 60.6                                      & 63.29                                     & 88.56                                     & 98.57                                     & \textbf{74.4}                                      & 59.45                                     & 63.66                                     & 71.16                                     & 54.14                                     & 47.41                                     & \textbf{59.16}                                     & 95                                        & 69.96                                                                         & 99.19                                     & 99.08                                     & 88.1                                       \\ 
\cline{2-19}
                                      & Mismatch Detection                        & 77.2                                      & 74.21                                     & 71.28                                     & 59.03                                     & 62.95                                     & \textbf{68.93}                                     & 74.16                                     & 70.17                                     & 64.25                                     & 26.86                                     & 18.69                                     & \textbf{50.83}                                     & 58.87                                     & 73.73                                                                        & 86.84                                     & 86.82                                     & 40.96                                      \\ 
\hhline{|~------------------|}
                                      & {\cellcolor[rgb]{0.871,0.871,0.871}}Ensemble Detection & {\cellcolor[rgb]{0.871,0.871,0.871}}98.9  & {\cellcolor[rgb]{0.871,0.871,0.871}}98.42 & {\cellcolor[rgb]{0.871,0.871,0.871}}97.07 & {\cellcolor[rgb]{0.871,0.871,0.871}}97.04 & {\cellcolor[rgb]{0.871,0.871,0.871}}99.78 & {\cellcolor[rgb]{0.871,0.871,0.871}}\textbf{98.24} & {\cellcolor[rgb]{0.871,0.871,0.871}}98.74 & {\cellcolor[rgb]{0.871,0.871,0.871}}97.35 & {\cellcolor[rgb]{0.871,0.871,0.871}}96.31 & {\cellcolor[rgb]{0.871,0.871,0.871}}68.09 & {\cellcolor[rgb]{0.871,0.871,0.871}}58.93 & {\cellcolor[rgb]{0.871,0.871,0.871}}\textbf{83.88} & {\cellcolor[rgb]{0.871,0.871,0.871}}99.3  & {\cellcolor[rgb]{0.871,0.871,0.871}}97.8  &  {\cellcolor[rgb]{0.871,0.871,0.871}}99.73 & {\cellcolor[rgb]{0.871,0.871,0.871}}99.8  & {\cellcolor[rgb]{0.871,0.871,0.871}}96.91  \\ 
\hline\hline
\multirow{4}{*}{\rotatebox[origin=c]{90}{CIFAR}}      & Selective Detection                       & 44.7                                      & 49.05                                     & 54.7                                      & 70.16                                     & 10.2                                      & \textbf{45.76}                                     & 33.74                                     & 33.25                                     & 30.6                                      & 22.68                                     & 18.68                                     & \textbf{27.79}                                     & 39.12                                     & 47.32                                                                        & 10.92                                     & 10.84                                     & 60.08                                      \\ 
\cline{2-19}
                                      & Confidence Detection                      & 72.57                                     & 81.45                                     & 87.35                                     & 99.03                                     & 99.99                                     & \textbf{88.08}                                     & 43.44                                     & 54.69                                     & 60.84                                     & 66.83                                     & 67.68                                     & \textbf{58.7}                                      & 86.53                                     & 81.74                                                                         & 92.78                                     & 93.05                                     & 87.96                                      \\ 
\cline{2-19}
                                      & Mismatch Detection                        & 27.62                                     & 36.19                                     & 42.58                                     & 42.67                                     & 58.52                                     & \textbf{41.52}                                     & 0                                         & 0                                         & 0.02                                      & 0.11                                      & 1.33                                      & \textbf{0.29}                                      & 37.79                                     & 43.54                                                                          & 39.15                                     & 38.92                                     & 54.35                                      \\ 
\hhline{|~------------------|}
                                      & {\cellcolor[rgb]{0.871,0.871,0.871}}Ensemble Detection& {\cellcolor[rgb]{0.871,0.871,0.871}}75.57 & {\cellcolor[rgb]{0.871,0.871,0.871}}82.82 & {\cellcolor[rgb]{0.871,0.871,0.871}}88.4  & {\cellcolor[rgb]{0.871,0.871,0.871}}99.36 & {\cellcolor[rgb]{0.871,0.871,0.871}}100   & {\cellcolor[rgb]{0.871,0.871,0.871}}\textbf{89.23} & {\cellcolor[rgb]{0.871,0.871,0.871}}57.74 & {\cellcolor[rgb]{0.871,0.871,0.871}}63.62 & {\cellcolor[rgb]{0.871,0.871,0.871}}66.66 & {\cellcolor[rgb]{0.871,0.871,0.871}}68.71 & {\cellcolor[rgb]{0.871,0.871,0.871}}68.74 & {\cellcolor[rgb]{0.871,0.871,0.871}}\textbf{65.09} & {\cellcolor[rgb]{0.871,0.871,0.871}}89.33 & {\cellcolor[rgb]{0.871,0.871,0.871}}87.38 &  {\cellcolor[rgb]{0.871,0.871,0.871}}93.21 & {\cellcolor[rgb]{0.871,0.871,0.871}}93.42 & {\cellcolor[rgb]{0.871,0.871,0.871}}92.76  \\
\hline
\end{tabular}
}
\end{table*}

\subsection{Performance on successful attacks only} \label{sec:suc_attacks}
Table \ref{tab:all_suc_attacks} shows the detection rate against the \acp{ae} that fooled the baseline \ac{dnn} classifier only under white-box and black-box scenarios.
For MNIST, in general, comparable results with the state-of-the-art detectors are achieved for all tested white and black boxes attacks ($>96.91\%$) except for the \ac{pgd} attacks (83.88\%). 
For both datasets, the impact of selective, confidence, and mismatch detection modules are obvious. The ability of the modules to detect the \acp{ae} is increasing when the amount of the perturbations is increasing. When the amount of the perturbations increases in a way that makes the adversarial samples feature space indistinguishable from the training dataset, the ability of these modules to detect the \acp{ae} decreases. Mismatch detection shows a high impact in the detection process of \acp{ae} except for \ac{pgd} attacks. Once the amount of crafted perturbation becomes high, the performance of mismatch detection decreases. That's because the detector classifiers' and the baseline \ac{dnn} classifier's behavior will be inconsistent for highly degraded inputs.

\begin{figure*}[t]
\begin{center}
\includegraphics[width=\linewidth, keepaspectratio]{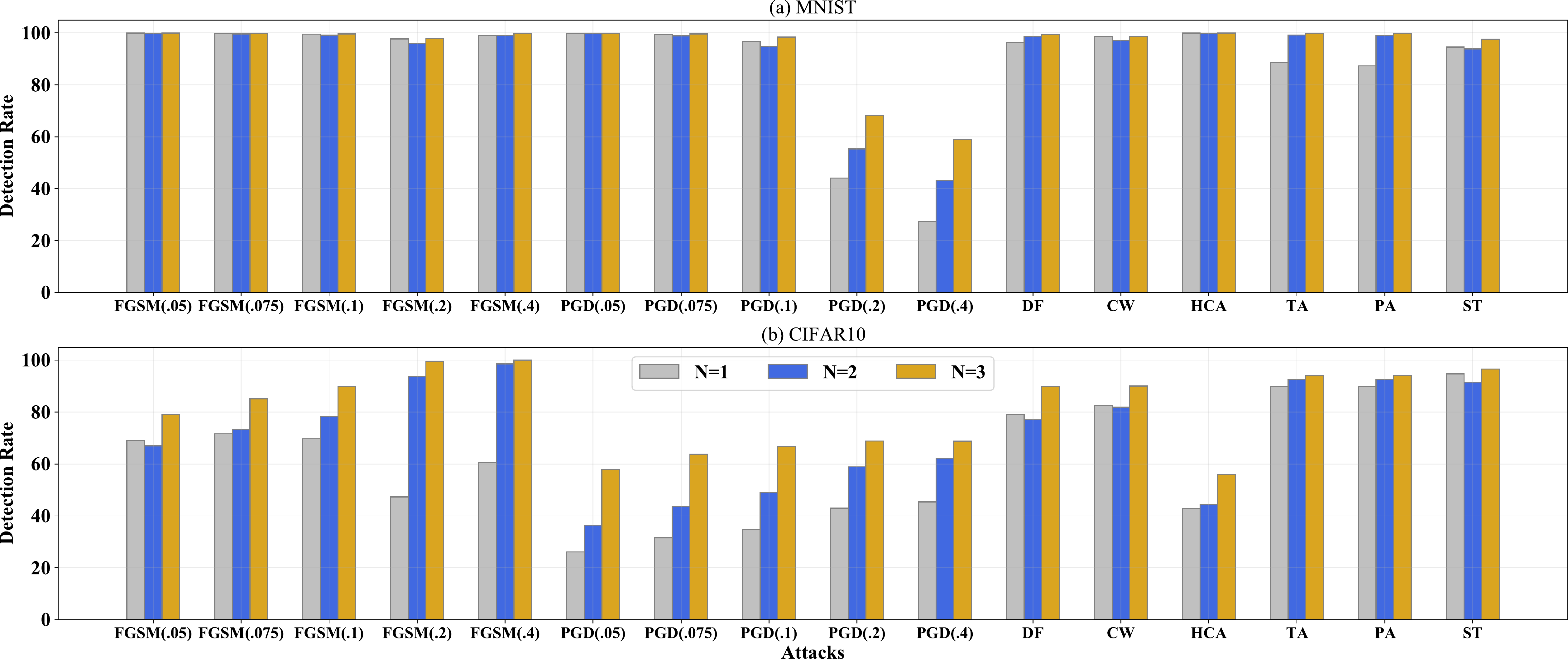}
\end{center}
\caption{Total model performance accuracy (\%) for black and white box scenarios on MINIST and CIFAR10 datasets at FP=10\% with different $N$ selective \acp{ae} classifiers settings}
\label{fig:last_n}
\end{figure*}

\subsection{Results with $N$ last layer(s) output(s)} \label{sec:dif_last_n}
Results shown in Figure \ref{fig:last_n} emphasize the conclusion in \cite{sotgiu2020deep} that recommends to use more than one layer from the last layers of the baseline \ac{dnn} classifier to be used in detection techniques. 
For MNIST dataset, the benefit of using more than one layer appears in detecting \ac{pgd} ($\epsilon=0.2 \text{, and 0.4}$), \ac{ta}, and \ac{pa} attacks, while it appears in all tested attacks on CIFAR10 dataset. It means that low-/and medium-level hidden layers hold features that will be triggered when small perturbations are added to input samples.    

\begin{table*}[t]
\centering
\caption{Ablation performance (\%) on white-box scenarios for MNIST dataset.}
\label{tab:mnist_ablation}
\resizebox{\textwidth}{!}{%
\begin{tabular}{|l||M{0.07\textwidth}||M{0.05\textwidth}|M{0.05\textwidth}|M{0.08\textwidth}|M{0.08\textwidth}|M{0.08\textwidth}|M{0.05\textwidth}|M{0.08\textwidth}|M{0.08\textwidth}|M{0.08\textwidth}|M{0.07\textwidth}|} 
\hline
Attack/Model & Baseline \ac{dnn} & {\cellcolor[rgb]{0.871,0.871,0.871}}NN    & Only noise & {\cellcolor[rgb]{0.871,0.871,0.871}}Only auto encoder & Only up/down sampling & {\cellcolor[rgb]{0.871,0.871,0.871}}Only bottleneck & No noise & {\cellcolor[rgb]{0.871,0.871,0.871}}No auto encoder & No up/down sampling & {\cellcolor[rgb]{0.871,0.871,0.871}}No bottleneck & Proposed  \\ 
\hline
\acs{fgsm}(0.05)   & 96.31      & {\cellcolor[rgb]{0.871,0.871,0.871}}\textbf{100}   & 99.97      & {\cellcolor[rgb]{0.871,0.871,0.871}}\textbf{100}               & \textbf{100}                   & {\cellcolor[rgb]{0.871,0.871,0.871}}99.92           & 99.95    & {\cellcolor[rgb]{0.871,0.871,0.871}}99.96           & 99.95               & {\cellcolor[rgb]{0.871,0.871,0.871}}99.97         & 99.96     \\ 
\hline
\acs{fgsm}(0.075)  & 92.93      & {\cellcolor[rgb]{0.871,0.871,0.871}}\textbf{99.98} & 99.94      & {\cellcolor[rgb]{0.871,0.871,0.871}}99.96             & 99.97                 & {\cellcolor[rgb]{0.871,0.871,0.871}}99.81           & 99.76    & {\cellcolor[rgb]{0.871,0.871,0.871}}99.9            & 99.8                & {\cellcolor[rgb]{0.871,0.871,0.871}}99.92         & 99.88     \\ 
\hline
\acs{fgsm}(0.1)    & 87.2       & {\cellcolor[rgb]{0.871,0.871,0.871}}99.9  & \textbf{99.93}      & {\cellcolor[rgb]{0.871,0.871,0.871}}99.9              & \textbf{99.93}                 & {\cellcolor[rgb]{0.871,0.871,0.871}}99.69           & 99.49    & {\cellcolor[rgb]{0.871,0.871,0.871}}99.79           & 99.51               & {\cellcolor[rgb]{0.871,0.871,0.871}}99.86         & 99.62     \\ 
\hline
\acs{fgsm}(0.2)    & 28.04      & {\cellcolor[rgb]{0.871,0.871,0.871}}\textbf{99.61} & 99.44      & {\cellcolor[rgb]{0.871,0.871,0.871}}98.97             & 98.78                 & {\cellcolor[rgb]{0.871,0.871,0.871}}96.71           & 96.37    & {\cellcolor[rgb]{0.871,0.871,0.871}}96.96           & 97.79               & {\cellcolor[rgb]{0.871,0.871,0.871}}98.79         & 97.86     \\ 
\hline
\acs{fgsm}(0.4)    & 7.91       & {\cellcolor[rgb]{0.871,0.871,0.871}}98.98 & 98.37      & {\cellcolor[rgb]{0.871,0.871,0.871}}98.65             & 96.14                 & {\cellcolor[rgb]{0.871,0.871,0.871}}97.85           & 93.73    & {\cellcolor[rgb]{0.871,0.871,0.871}}\textbf{99.87}           & 98.81               & {\cellcolor[rgb]{0.871,0.871,0.871}}97.01         & 99.8      \\ 
\hline
\acs{pgd}(0.05)    & 95.18      & {\cellcolor[rgb]{0.871,0.871,0.871}}\textbf{100}   & 99.95      & {\cellcolor[rgb]{0.871,0.871,0.871}}99.97             & 99.98                 & {\cellcolor[rgb]{0.871,0.871,0.871}}99.88           & 99.87    & {\cellcolor[rgb]{0.871,0.871,0.871}}99.93           & 99.91               & {\cellcolor[rgb]{0.871,0.871,0.871}}99.95         & 99.94     \\ 
\hline
\acs{pgd}(0.075)   & 85.84      & {\cellcolor[rgb]{0.871,0.871,0.871}}99.83 & \textbf{99.86}      & {\cellcolor[rgb]{0.871,0.871,0.871}}99.84             & \textbf{99.86}                 & {\cellcolor[rgb]{0.871,0.871,0.871}}99.58           & 99.43    & {\cellcolor[rgb]{0.871,0.871,0.871}}99.73           & 99.57               & {\cellcolor[rgb]{0.871,0.871,0.871}}99.76         & 99.63     \\ 
\hline
\acs{pgd}(0.1)     & 56.91      & {\cellcolor[rgb]{0.871,0.871,0.871}}\textbf{99.35} & 99.2       & {\cellcolor[rgb]{0.871,0.871,0.871}}99.06             & 99.09                 & {\cellcolor[rgb]{0.871,0.871,0.871}}98.09           & 97.65    & {\cellcolor[rgb]{0.871,0.871,0.871}}98.43           & 97.7                & {\cellcolor[rgb]{0.871,0.871,0.871}}98.92         & 98.4      \\ 
\hline
\acs{pgd}(0.2)     & 0          & {\cellcolor[rgb]{0.871,0.871,0.871}}60.06 & 59.33      & {\cellcolor[rgb]{0.871,0.871,0.871}}66.24             & 53.73                 & {\cellcolor[rgb]{0.871,0.871,0.871}}66.18           & 57.9     & {\cellcolor[rgb]{0.871,0.871,0.871}}61.39           & 63.44               & {\cellcolor[rgb]{0.871,0.871,0.871}}63.93         & \textbf{68.09}     \\ 
\hline
\acs{pgd}(0.4)     & 0          & {\cellcolor[rgb]{0.871,0.871,0.871}}43.8  & 42.61      & {\cellcolor[rgb]{0.871,0.871,0.871}}52.57             & 40.97                 & {\cellcolor[rgb]{0.871,0.871,0.871}}46.11           & 45.11    & {\cellcolor[rgb]{0.871,0.871,0.871}}43.39           & 46.57               & {\cellcolor[rgb]{0.871,0.871,0.871}}47.71         & \textbf{58.93}     \\ 
\hline
\acs{df}           & 4.68       & {\cellcolor[rgb]{0.871,0.871,0.871}}97.71 & 97.35      & {\cellcolor[rgb]{0.871,0.871,0.871}}98.44             & 95.15                 & {\cellcolor[rgb]{0.871,0.871,0.871}}98.51           & 96.88    & {\cellcolor[rgb]{0.871,0.871,0.871}}99              & 98.83               & {\cellcolor[rgb]{0.871,0.871,0.871}}98.62         & \textbf{99.33}     \\ 
\hline
\acs{cw}           & 38.97      & {\cellcolor[rgb]{0.871,0.871,0.871}}\textbf{99.71} & 99.66      & {\cellcolor[rgb]{0.871,0.871,0.871}}99.52             & 99.6                  & {\cellcolor[rgb]{0.871,0.871,0.871}}98.36           & 97.73    & {\cellcolor[rgb]{0.871,0.871,0.871}}98.79           & 98.57               & {\cellcolor[rgb]{0.871,0.871,0.871}}99.19         & 98.65     \\ 
\hline
\end{tabular}
}
\end{table*}

\begin{table*}[t]
\centering
\caption{Ablation performance (\%) on white-box scenarios for CIFAR10 dataset.}
\label{tab:cifar_ablation}
\resizebox{\textwidth}{!}{%
\begin{tabular}{|l||M{0.07\textwidth}||M{0.05\textwidth}|M{0.05\textwidth}|M{0.08\textwidth}|M{0.08\textwidth}|M{0.08\textwidth}|M{0.05\textwidth}|M{0.08\textwidth}|M{0.08\textwidth}|M{0.08\textwidth}|M{0.07\textwidth}|} 
\hline
Attack/Model & Baseline \ac{dnn} & {\cellcolor[rgb]{0.871,0.871,0.871}}NN    & Only noise & {\cellcolor[rgb]{0.871,0.871,0.871}}Only auto encoder & Only up/down sampling & {\cellcolor[rgb]{0.871,0.871,0.871}}Only bottleneck & No noise & {\cellcolor[rgb]{0.871,0.871,0.871}}No auto encoder & No up/down sampling & {\cellcolor[rgb]{0.871,0.871,0.871}}No bottleneck & Proposed  \\ 
\hline
\acs{fgsm}(0.05)   & 14.09      & {\cellcolor[rgb]{0.871,0.871,0.871}}\textbf{80.02} & 79.55      & {\cellcolor[rgb]{0.871,0.871,0.871}}73                & 80.01                 & {\cellcolor[rgb]{0.871,0.871,0.871}}79.68           & 78.62    & {\cellcolor[rgb]{0.871,0.871,0.871}}75.81           & 75.7                & {\cellcolor[rgb]{0.871,0.871,0.871}}78.55         & 79.01     \\ 
\hline
\acs{fgsm}(0.075)  & 13.44      & {\cellcolor[rgb]{0.871,0.871,0.871}}\textbf{87.9}  & 85.21      & {\cellcolor[rgb]{0.871,0.871,0.871}}80.44             & 85.84                 & {\cellcolor[rgb]{0.871,0.871,0.871}}86.61           & 86.2     & {\cellcolor[rgb]{0.871,0.871,0.871}}84.77           & 81.3                & {\cellcolor[rgb]{0.871,0.871,0.871}}85            & 85.12     \\ 
\hline
\acs{fgsm}(0.1)    & 12.25      & {\cellcolor[rgb]{0.871,0.871,0.871}}\textbf{91.37} & 87.26      & {\cellcolor[rgb]{0.871,0.871,0.871}}84.64             & 87.71                 & {\cellcolor[rgb]{0.871,0.871,0.871}}91              & 91.27    & {\cellcolor[rgb]{0.871,0.871,0.871}}90.75           & 85.56               & {\cellcolor[rgb]{0.871,0.871,0.871}}88.26         & 89.81     \\ 
\hline
\acs{fgsm}(0.2)    & 10.5       & {\cellcolor[rgb]{0.871,0.871,0.871}}85.52 & 78.89      & {\cellcolor[rgb]{0.871,0.871,0.871}}95.13             & 78.73                 & {\cellcolor[rgb]{0.871,0.871,0.871}}97.79           & \textbf{99.48}    & {\cellcolor[rgb]{0.871,0.871,0.871}}98.19           & 95.69               & {\cellcolor[rgb]{0.871,0.871,0.871}}88.92         & 99.43     \\ 
\hline
\acs{fgsm}(0.4)    & 9.75       & {\cellcolor[rgb]{0.871,0.871,0.871}}84.39 & 84.94      & {\cellcolor[rgb]{0.871,0.871,0.871}}99.71             & 79.32                 & {\cellcolor[rgb]{0.871,0.871,0.871}}\textbf{100}             & \textbf{100}      & {\cellcolor[rgb]{0.871,0.871,0.871}}99.98           & \textbf{100}                 & {\cellcolor[rgb]{0.871,0.871,0.871}}99.85         & \textbf{100}       \\ 
\hline
\acs{pgd}(0.05)    & 0.43       & {\cellcolor[rgb]{0.871,0.871,0.871}}0.42  & 0.43       & {\cellcolor[rgb]{0.871,0.871,0.871}}4.21              & 0.42                  & {\cellcolor[rgb]{0.871,0.871,0.871}}1.08            & 36.61    & {\cellcolor[rgb]{0.871,0.871,0.871}}7.8             & 48.34               & {\cellcolor[rgb]{0.871,0.871,0.871}}12.58         & \textbf{57.91}     \\ 
\hline
\acs{pgd}(0.075)   & 0.28       & {\cellcolor[rgb]{0.871,0.871,0.871}}0.27  & 0.28       & {\cellcolor[rgb]{0.871,0.871,0.871}}11.64             & 0.27                  & {\cellcolor[rgb]{0.871,0.871,0.871}}2.76            & 47.19    & {\cellcolor[rgb]{0.871,0.871,0.871}}9.65            & 58.32               & {\cellcolor[rgb]{0.871,0.871,0.871}}12.67         & \textbf{63.72}     \\ 
\hline
\acs{pgd}(0.1)     & 0.22       & {\cellcolor[rgb]{0.871,0.871,0.871}}0.22  & 0.22       & {\cellcolor[rgb]{0.871,0.871,0.871}}17.72             & 0.22                  & {\cellcolor[rgb]{0.871,0.871,0.871}}5.22            & 52.68    & {\cellcolor[rgb]{0.871,0.871,0.871}}12.09           & 61.49               & {\cellcolor[rgb]{0.871,0.871,0.871}}13.42         & \textbf{66.74}     \\ 
\hline
\acs{pgd}(0.2)     & 0.16       & {\cellcolor[rgb]{0.871,0.871,0.871}}0.15  & 0.15       & {\cellcolor[rgb]{0.871,0.871,0.871}}27.83             & 0.16                  & {\cellcolor[rgb]{0.871,0.871,0.871}}14.7            & 58.23    & {\cellcolor[rgb]{0.871,0.871,0.871}}18.86           & 64.31               & {\cellcolor[rgb]{0.871,0.871,0.871}}16.61         & \textbf{68.83}     \\ 
\hline
\acs{pgd}(0.4)     & 0.17       & {\cellcolor[rgb]{0.871,0.871,0.871}}0.17  & 0.17       & {\cellcolor[rgb]{0.871,0.871,0.871}}33.74             & 0.16                  & {\cellcolor[rgb]{0.871,0.871,0.871}}22.75           & 60.23    & {\cellcolor[rgb]{0.871,0.871,0.871}}23.9            & 65.4                & {\cellcolor[rgb]{0.871,0.871,0.871}}18.21         & \textbf{68.78}     \\ 
\hline
\acs{df}           & 4.79       & {\cellcolor[rgb]{0.871,0.871,0.871}}83.67 & 84.84      & {\cellcolor[rgb]{0.871,0.871,0.871}}83.71             & 82.7                  & {\cellcolor[rgb]{0.871,0.871,0.871}}89.11           & 88.32    & {\cellcolor[rgb]{0.871,0.871,0.871}}87.64           & 87.96               & {\cellcolor[rgb]{0.871,0.871,0.871}}88.09         & \textbf{89.8 }     \\ 
\hline
\acs{cw}           & 20.95      & {\cellcolor[rgb]{0.871,0.871,0.871}}88.98 & 89.05      & {\cellcolor[rgb]{0.871,0.871,0.871}}87.03             & 88.33                 & {\cellcolor[rgb]{0.871,0.871,0.871}}89.46           & 88.63    & {\cellcolor[rgb]{0.871,0.871,0.871}}87.7            & 88.14               & {\cellcolor[rgb]{0.871,0.871,0.871}}89.26         & \textbf{90.02}     \\ 
\hline
\end{tabular}
}
\end{table*}

\subsection{Ablation study} \label{sec:ablation_study} 
In this section, we emphasize the advantages of \ac{sfad}'s feature processing components including noise, autoencoder, up/down sampling, and bottleneck blocks. Table \ref{tab:mnist_ablation} and Table \ref{tab:cifar_ablation} show the performance results for each block once when it is present alone and another time when it is absent for MNIST and CIFAR10 datasets. In all settings, the selectiveNet is present in the selective \ac{ae} classifiers and in the selective knowledge transfer classifier. 

\textbf{Only \ac{nn}.} When all processing blocks are absent, the MNIST results show the ability to detect \ac{fgsm}, \acs{pgd} of small $\epsilon$ values, and \ac{cw} attacks slightly better than the proposed approach. While the proposed approach yields better results for \ac{df} and \ac{pgd} of high $\epsilon$ values. Since CIFAR10 dataset is different from MNIST and has different characteristics, the only \ac{nn} component did not yield better results against \ac{fgsm} of high $\epsilon$ values, \ac{pgd}, \ac{cw}, and DF attacks.

\textbf{Noise.} When only the noise block is used, the model achieves comparable results to \ac{sfad} except against \ac{pgd} attacks. When we remove the noise block, the performance of \ac{sfad} is reduced especially against \ac{pgd} attacks for MNIST and CIFAR10 datasets. The noise block helps the detector to better distinguish the feature space of clean input images from those features of \acp{ae}.

\textbf{Autoencoder.} Autoencoder block shows a substantial impact in the proposed approach. As discussed in Section \ref{sec:feature_processing_blocks}, if the autoencoder couldn't reconstruct its input, different feature space might be generated for the input signal which let \ac{sfad} able to detect the \acp{ae}. For MNIST dataset, the autoencoder block enhanced the performance results compared to only \ac{nn} model against \ac{pgd} of higher $\epsilon$ values, while the performance is reduced when the autoencoder block is removed from the proposed approach. 
On the other hand, for CIFAR10 dataset, when only the autoencoder is present, the performance results are much better against \ac{fgsm} of high $\epsilon$ values, \ac{pgd}, \ac{cw}, and \ac{df} attacks when it is compared to only \ac{nn}. The performance is reduced when it is removed from the proposed approach against \ac{pgd} attacks. 

\textbf{Up/down-sampling.} Unlike other processing blocks, up/ down sampling block yields less performance results against \ac{fgsm} attacks and yields comparable results against other attacks compared to only \ac{nn} model. That's because the up/ down-sampling restores the global information of the input signal by the average pooling process. On the other hand, removing the sampling block from the proposed approach reduces the performance results especially for the CIFAR10 dataset.

\textbf{Bottleneck.} Like autoencoder block, the bottleneck block shows its ability to distinguish input signal characteristics especially in the proposed shallow classifiers (the selective \acp{ae} classifiers). Compared to only \ac{nn} model, the only bottleneck model enhanced the performance results against \ac{fgsm} of high $\epsilon$ values, \ac{pgd}, \ac{cw}, and \ac{df} attacks for CIFAR10 dataset and enhanced the performance results against \ac{pgd} of high $\epsilon$ values attacks for MNIST. Besides, the performance of the proposed approach is significantly decreased for CIFAR10 dataset when the bottleneck block is removed.

\begin{figure*}[t]
\begin{center}
\includegraphics[width=\linewidth, keepaspectratio]{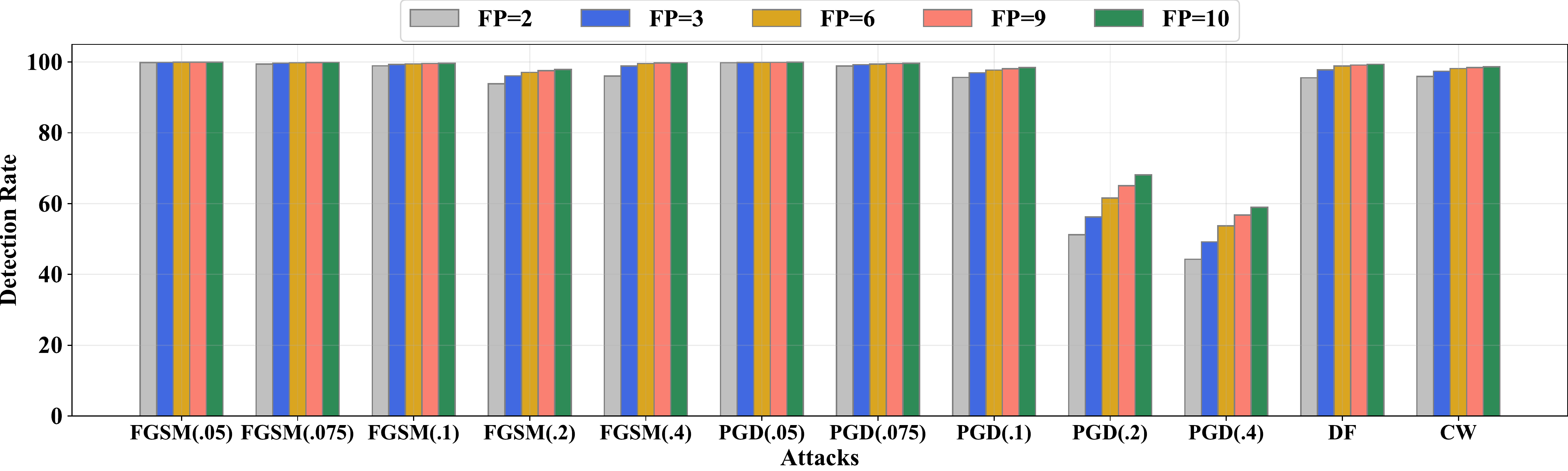}
\end{center}
\caption{Performance comparisons between different False Positive (FP) rates and FP=10\% of \ac{sfad} for white-box attacks on MNIST dataset.}
\label{fig:fprs}
\end{figure*}

\subsection{Performance with different rejection rates (False positive (FP))} \label{sec:rejection_rates}
In this subsection we show the performance results of the proposed approach when thresholds are set to reject less than 10\% for MNIST as shown in Figure \ref{fig:fprs}. 
Results show that an acceptable performance can be achieved if the thresholds are set to less than 10\%. For instance, when the false positive rate is set to be 2\%, results against \ac{pgd} ($\epsilon=0.2,\text{, and }0.4$) attacks are significantly decreased because of the selective detection. In all other tested attacks, the difference is up to 4\% and 1.76\% when FP=2\% and 3\%, respectively.

\begin{table*}[t]
\centering
\caption{Detection accuracies for the state-of-the-art detectors against white-box and black-box attacks. Top 3 are colored with \textcolor{red}{red}, \textcolor{blue}{blue}, and \textcolor{customgreen}{green} respectively.}

\label{tab:comparison_mnist_cifar10}
\begin{tabular}{c|c|c|c|c|c|c|c|c|c|c|}
\hline
\multirow{3}{*}{Dataset} & \multirow{3}{*}{Detector} & \multicolumn{9}{c|}{Attacks}                                                                                                                                    \\ \cline{3-11} 
                         &                           & \multirow{2}{*}{FPR} & \multicolumn{5}{c|}{White box}                                                        & \multicolumn{3}{c|}{Black box}                   \\ \cline{4-11} 
                         &                           &                      & FGSM            & PGD             & CW             & HCA             & DF             & SA             & HSJA           & STA            \\ \hline\hline
\multirow{7}{*}{MNIST}   & \ac{kd}+\ac{bu}$^\dagger$                       & 1.76                 & 62.69           & 52.52           & 42.77          & 59.26           & 48.97           & 53.53         & 61.82          & 47.94 \\ \cline{2-11}
                          & LID$^\dagger$                       & 0.81                 & 77.46           & 77.03           & 64.43          & 84.51           & 93.3           & 42.78          & 61.52          & \textcolor{customgreen}{\textbf{93.81}} \\ \cline{2-11} 
                         & FS$^\ast$                        & 5.27                 & 97.96           & \textcolor{customgreen}{\textbf{97.19}}  & \textcolor{blue}{\textbf{98.41}} & \textcolor{customgreen}{\textbf{99.99}}  & 66.96          & \textcolor{red}{\textbf{99.96}} & \textcolor{blue}{\textbf{99.98}} & 77.49          \\ \cline{2-11} 
                         & MagNet$^\ast$                     & 0.20                 & \textcolor{red}{\textbf{100.00}} & \textcolor{red}{\textbf{100.00}} & 40.56          & \textcolor{red}{\textbf{100.00}} & \textcolor{customgreen}{\textbf{96.99}} & \textcolor{blue}{\textbf{99.93}} & 98.32          & 1.61           \\ \cline{2-11} 
                         & NIC$^\ast$                        & 10.12                & \textcolor{blue}{\textbf{100.00}} & \textcolor{blue}{\textbf{100.00}} & \textcolor{red}{\textbf{100}}   & \textcolor{blue}{\textbf{100.00}} & \textcolor{red}{\textbf{100}}   & \textcolor{customgreen}{\textbf{99.68}} & \textcolor{red}{\textbf{100}}   & \textcolor{red}{\textbf{99.83}} \\ \cline{2-11} 
                         & DNR$^\ast$                        & 10.01                & 79.67           & 59.21           & 57.98          & 89.90           & 95.6           & 81.27          & 59.98          & 88             \\ \cline{2-11} \cline{2-11} 
                         & SFAD$^\ast$  (ours)                     & 10.79                & \textcolor{customgreen}{\textbf{98.66}}  & 81.83           & \textcolor{customgreen}{\textbf{98.24}} & 99.36           & \textcolor{blue}{\textbf{99.58}} & 98.85          & \textcolor{customgreen}{\textbf{99.91}} & \textcolor{blue}{\textbf{97.61}} \\ \hline\hline

\multirow{7}{*}{CIFAR10} & \ac{kd}+\ac{bu}$^\dagger$                         & 3.57                 & 34.13           & \textcolor{customgreen}{\textbf{96.08}}            & 21.12          & 33.49  & 54.02          & 0          & 28.03 & 44.15          \\ \cline{2-11} & LID$^\dagger$                       & 7.30                 & \textcolor{blue}{\textbf{76.15}}  & \textcolor{blue}{\textbf{96.81}}  & \textcolor{blue}{\textbf{64.52}} & \textcolor{customgreen}{\textbf{55.03}}  & \textcolor{customgreen}{\textbf{63.57}} & \textcolor{customgreen}{\textbf{85.76}} & \textcolor{blue}{\textbf{88.34}} & \textcolor{red}{\textbf{94.23}} \\ \cline{2-11} 
                         & FS$^\ast$                         & 5.07                 & 32.50           & 4.20            & 56.18          & \textcolor{blue}{\textbf{56.31}}  & 39.18          & 17.82          & \textcolor{customgreen}{\textbf{84.16}} & 22.46          \\ \cline{2-11} 
                         & MagNet$^\ast$                     & 0.77                 & 34.61           & 0.62            & 13.23          & 0.53            & 57.33          & \textcolor{red}{\textbf{94.04}} & 0.58           & 2.32           \\ \cline{2-11} 
                         & NIC$^\ast$                       & 10.08                & \textcolor{customgreen}{\textbf{63.15}}  & \textcolor{red}{\textbf{100.00}} & \textcolor{customgreen}{\textbf{61.68}} & \textcolor{red}{\textbf{73.29}}  & \textcolor{blue}{\textbf{84.91}} & 61.88          & 67.53          & 48.77          \\ \cline{2-11} 
                         & DNR$^\ast$                        & 10.01                & 30.23           & 18.23           & 44.15          & 29.85           & 30.2           & 52.86          & 38.81          & \textcolor{customgreen}{\textbf{56.2}}  \\ \cline{2-11} \cline{2-11}
                         & SFAD$^\ast$  (ours)                      & 10.90                & \textcolor{red}{\textbf{80.14}}  & 41.2  & \textcolor{red}{\textbf{87.68}} & 45.85           & \textcolor{red}{\textbf{89.57}} & \textcolor{blue}{\textbf{93.91}} & \textcolor{red}{\textbf{95.57}} & \textcolor{blue}{\textbf{92.9}}  \\ \hline
                         \multicolumn{11}{c}{$\dagger$ Supervised detector, $^\ast$ Unsupervised detector}\\
\end{tabular}
\end{table*}

\subsection{Robustness against high confidence attack}
In \cite{athalye2018obfuscated}, ten defenses and detectors were broken using \ac{bpda}, \ac{eot}, and \acf{hca}. \ac{bpda}, and \ac{eot} are appropriate for defense techniques, while \ac{hca} is used to fail detectors. \ac{hca} is a variant of \ac{cw} attack and generates adversarial examples with high confidence level. In \cite{athalye2018obfuscated}, \ac{lid} were broken using \ac{hca}. In his experiment, we generate \acp{ae} using \ac{hca} with $\epsilon=0.3125$ for MNIST and $\epsilon=0.031$ for CIFAR10. The results show that \ac{sfad} is fully robust on MNIST against \ac{hca} and partially robust (57.76\%) on CIFAR10. Our analysis finds that the confidence and selective detection methods are effective to detect \acp{ae}. In case the confidence level of the attack is increased, \ac{sfad} can be fine-tuned by selecting the proper layer outputs to build the selective \ac{ae} classifiers. 

All the experiments that are conducted in this work are tested under zero knowledge of the detector. We assume that the adversary's work is very hard for building an adaptive attack to fool \ac{sfad} since it ensembles three detection methods. Despite that, \ac{sfad} performance will drop when the adversary is able to craft customized perturbations to fool both the baseline classifier and the ensemble detector. 

\subsection{Comparisons with state-of-the-art detectors}\label{sec:compare_sota} 
In this subsection we build a comparison with different types of supervised and unsupervised detectors using the detectors benchmark\footnote{The source code is available in \url{https://github.com/aldahdooh/detectors_review}} and the results are shown in Table \ref{tab:comparison_mnist_cifar10}. We compare the average \ac{fgsm} and \ac{pgd} results. For fair comparisons, $\epsilon$ values of 0.125, 0.25, and 0.3125 are set for MNIST dataset, while for CIFAR10 are set to 0.03 and 0.06. Moreover, the supervised detectors are trained and tested separately against each adversarial attack algorithm. As discussed in Section \ref{sec:rejection_rates}, rejection/false positive rates of \ac{sfad} can be decreased with small compromise in the performance. For RAID \cite{eniser2020raid} method, we compare with the results that are reported in the original paper due to publicly unavailable code.

\textbf{\acs{kd}+\acs{bu} \cite{feinman2017detecting}}. \acs{kd}+\acs{bu} detector is a combination of kernel density and Bayesian uncertainty based classifiers. For both datasets, the results show that \ac{sfad} outperforms \acs{kd}+\acs{bu} detector against all tested attacks except for \ac{pgd} attacks in CIFAR10 dataset. In fact, \acs{kd}+\acs{bu} needs not noisy clean and adversarial images to accurately train the detector to identify the boundaries between clean and adversarial inputs.

\textbf{\acs{lid} \cite{ma2018characterizing}}. \ac{sfad} outperforms \ac{lid} in both datasets and the tested attacks except for \ac{pgd} attacks on CIFAR10. \ac{lid} achieves better false positive rate compared to \ac{sfad} but it fails against \acl{hca} as reported in  \cite{carlini2017adversarial}. When \ac{lid} is trained for the \ac{hca} attacks, it achieves better results than in \cite{carlini2017adversarial}. Our approach provides full and partial robustness against \ac{hca} for MNIST and CIFAR10, respectively. Similar to \acs{kd}+\acs{bu}, \ac{lid} needs not noisy clean and adversarial images to accurately train the detector to identify the boundaries between clean and adversarial inputs.

\textbf{RAID \cite{eniser2020raid}}. For MNIST datasets, RAID achieves higher detection rate for \ac{pgd} attacks ($\epsilon=0.3$) and higher detection rate against \ac{fgsm} and \ac{pgd} attacks for CIFAR10 while our approach improved the performance against \ac{cw} and \ac{df} attacks. Besides, RAID has a better false positive rate for MNIST only. RAID trains clean and adversarial inputs to identify differences in neuron activation between clean and adversarial samples. Hence, it requires a huge knowledge of attacks and its variants to enhance its performance.

\textbf{FS \cite{xu2017feature}}. As stated in \cite{xu2017feature}, FS requires high quality squeezers for different baseline networks and it was shown that FS is not performing well against tested attacks on CIFAR10 dataset, while our approach generalizes better than FS at the expense of higher false positive rate.

\textbf{MagNet \cite{meng2017magnet}}. Results reported on Table \ref{tab:comparison_mnist_cifar10} is for the detection process of MagNet and defense process of MagNet is not considered. For MNIST, comparable results are achieved by our approach except for \ac{cw} and \ac{st} attacks where \ac{sfad} achieves better performance. 
For CIFAR, our approach outperforms MagNet against the tested attacks. Since MagNet is a denoiser-based detector, it is not guaranteed that the denoisers will remove all the noise and have highly denoised inputs that respect the target threshold. This applies specifically to $L_0$ and $L_2$ attacks. On the contrary, our approach relies on confidence value changes that the \acp{ae} will cause which makes our approach able to identify \acp{ae}. Although MagNet yields to a less false positive rate, it was shown in \cite{carlini2017magnet} that MagNet can be broken by different strategies. 

\textbf{\acs{nic} \cite{ma2019nic}}. \ac{nic} is the state-of-the-art detector that achieves better performance, in general, against white box attacks compared to other detectors, while our approach achieves better performance against tested black box attacks. Unlike the proposed approach, other works reported that the \ac{nic}'s baseline detectors are not consistent \cite{bulusu2020anomalous}, increase the  model parameters overhead \cite{lust2020gran}, are time consuming \cite{ma2019nic}, and have latency in the inference time \cite{gao2020backdoor}.

\textbf{\ac{dnr} \cite{sotgiu2020deep}}. \ac{dnr} adopted confidence-based detectors and is close to our approach, but we include the feature processing and selective modules components. The reported results show that our approach outperforms \ac{dnr} at the same false positive rates for MNIST and CIFAR10 datasets.

\textbf{Other performance comparison: } \ac{sfad} has middle complexity level due to classifiers training times, and has no inference time latency, but it has a compromise on overhead due to classifiers parameters saving. Compared to other detectors, \ac{sfad} introduces shallow networks hence, compared to \ac{nic}, \ac{dnr}, and \ac{lid}, our detector has much less complexity. Besides, it works in parallel to the baseline classifier and no latency is provided compared to \ac{fs} and \ac{nic}. Finally, like \ac{nic} and \ac{dnr}, \ac{sfad} has to pay a little price in terms of overhead compared to MagNet, \ac{fs} and \ac{lid}.

\section{Conclusion}
In this work, we have proposed a novel unsupervised and ensemble mechanism, namely \ac{sfad}, to detect adversarial attacks. \ac{sfad} handled the $N$-last layers outputs of the baseline \ac{dnn} classifier to identify \acp{ae}. It built $N$ selective \acp{ae} classifiers that each took one layer output of the baseline classifier as input and then processed the input using autoencoder, up/down sampling, bottleneck, and additive noise blocks. Then, these feature-based classifiers were optimized in the SelectiveNet model to estimate the model's uncertainties and confidences. The confidence values of these classifiers were then distilled as input to the selective knowledge transfer classifier to build the last classifier. Selective and confidence thresholds were set to identify the adversarial inputs. Selective, confidence, and mismatch modules are jointly working to enhance the detection accuracy. We showed that the model is consistent and is able to detect tested attacks. Moreover, the model is robust in different attack scenarios; white, black, and gray boxes attacks. This robustness, with the advantage that the model does not require any knowledge of adversarial attacks, will lead to better generalization. The main limitation of the model is that the best combination of $N$ needs to be identified to enhance the detection accuracy and to reduce the false positive rate.

\section*{Acknowledgement}
The project is funded by both R\'egion Bretagne (Brittany region), France, and direction g\'en\'erale de l'armement (DGA).


{
\bibliographystyle{unsrt}
\bibliography{refs}
}
\vfill
\pagebreak

\end{document}